\documentclass[fleqn,usenatbib]{mnras}
\usepackage{newtxtext,newtxmath}
\usepackage[T1]{fontenc}
\usepackage{ae,aecompl}

\usepackage{graphicx}	
\usepackage[all]{hypcap} 
\usepackage{upgreek}
\usepackage{subcaption}
\usepackage{comment}
\usepackage{enumitem}
\usepackage[normalem]{ulem} 

\renewcommand{\eqref}[1]{equation\ (\ref{#1})}
\newcommand{\bba}{$^{\scriptstyle 3\mathrm{D}}$B{\sc arolo}}
\newcommand{\hi}{\ifmmode{\rm HI}\else{H\/{\sc i}}\fi}
\newcommand{\ha}{\ifmmode{\rm H\upalpha}\else{H$\upalpha$}\fi} 
\newcommand{\nii}{$\mathrm{[N\,\textsc{ii}]}$}
\newcommand{\de}{\ifmmode{^\circ}\else{$^\circ$}\fi} 
\newcommand{\vlos}{\ifmmode{V_\mathrm{los}}\else{$V_\mathrm{los}$}\fi}
\newcommand{\vsys}{\ifmmode{V_\mathrm{sys}}\else{$V_\mathrm{sys}$}\fi}
\newcommand{\vrot}{\ifmmode{V_\mathrm{rot}}\else{$V_\mathrm{rot}$}\fi}
\newcommand{\vflat}{\ifmmode{V_\mathrm{flat}}\else{$V_\mathrm{flat}$}\fi}
\newcommand{\vdisp}{\ifmmode{\sigma_\mathrm{gas}}\else{$\sigma_\mathrm{gas}$}\fi}
\newcommand{\sigmastar}{\ifmmode{\Sigma_\mathrm{\star}}\else{$\Sigma_\mathrm{\star}$}\fi}
\newcommand{\jstar}{\ifmmode{j_{\star}}\else{$j_{\star}$}\fi}
\newcommand{\mstar}{\ifmmode{M_{\star}}\else{$M_{\star}$}\fi}
\newcommand{\mo}{{\rm M}_\odot}
\newcommand{\moyr}{{\rm M}_\odot \, {\rm yr^{-1}}}

\newcommand{\kms} {\,{\rm km\,s}^{-1}}

\newcommand{\ang}{\,\text{\rm \AA}}
\newcommand{\Mh}{M_{\rm h}}
\newcommand{\Vh}{V_{\rm h}}
\newcommand{\jh}{j_{\rm h}}
\newcommand{\galnum}{43}

\title[Scaling relations of massive spiral galaxies]{Rotation curves and scaling relations of extremely massive spiral galaxies}

\author[Di Teodoro et al.]{Enrico\ M.\ Di Teodoro$^{1,2}$\thanks{E-mail: editeodoro@jhu.edu},
Lorenzo Posti$^{3}$, 
Patrick M.\ Ogle$^{2}$,
S.\ Michael Fall$^{2}$
\newauthor
and Thomas Jarrett$^{4}$ \\
$^{1}$Department of Physics \& Astronomy, Johns Hopkins University, Baltimore, MD 21218, USA\\
$^{2}$Space Telescope Science Institute, 3700 San Martin Drive, Baltimore, MD 21218, USA\\
$^{3}$Universit\'{e} de Strasbourg, CNRS UMR 7550, Observatoire astronomique de Strasbourg, 11 rue de l’Universit\'{e}, 67000 Strasbourg, France\\
$^{4}$University of Cape Town, Cape Town, South Africa\\
}

\begin{document}
\label{firstpage}
\pagerange{\pageref{firstpage}--\pageref{lastpage}}
\maketitle

\begin{abstract}
We study the kinematics and scaling relations of a sample of \galnum\ giant spiral galaxies that have stellar masses exceeding $10^{11} \, \mo$ and optical discs up to 80 kpc in radius. 
We use a hybrid 3D-1D approach to fit 3D kinematic models to long-slit observations of the \ha-\nii\ emission lines and we obtain robust rotation curves of these massive systems.
We find that all galaxies in our sample seem to reach a flat part of the rotation curve within the outermost optical radius. 
We use the derived kinematics to study the high-mass end of the two most important scaling relations for spiral galaxies: the stellar/baryonic mass Tully-Fisher relation and the Fall (mass-angular momentum) relation.
All galaxies in our sample, with the possible exception of the two fastest rotators, lie comfortably on both these scaling relations determined at lower masses, without any evident break or bend at the high-mass regime.
When we combine our high-mass sample with lower-mass data from the Spitzer Photometry \& Accurate Rotation Curves catalog, we find a slope of $\alpha=4.25\pm0.19$ for the stellar Tully-Fisher relation and a slope of $\gamma=0.64\pm0.11$ for the Fall relation. 
Our results indicate that most, if not all, of these rare, giant spiral galaxies are scaled up versions of less massive discs and that spiral galaxies are a self-similar population of objects up to the very high-mass end. 
\end{abstract}




\section{Introduction}
\label{sec:intro}

Disc galaxies follow tight, and approximately featureless, scaling relations between some of their basic global properties, such as mass, velocity, and angular momentum.
The most important scaling laws are arguably the \citet{Tully+1977} relation, between the stellar mass or luminosity of a galaxy and its rotation velocity $V$, and the \citet{Fall+1983} relation, between stellar mass $\mstar$ and specific angular momentum $\jstar\equiv J_\star / \mstar$.
The stellar Tully-Fisher is a power law $\mstar \propto V^\alpha$ with $\alpha\sim 4-5$ depending on the prescriptions used to estimate stellar masses \citep{McGaugh+2015, Ponomareva+2018} and depending on how the rotation velocities are defined \citep[][]{Verheijen+2001,Lelli+2019}.
When the baryonic mass $M_{\rm bar}$ (stars + cold gas) is used instead of the stellar mass, the baryonic Tully-Fisher relation \citep{McGaugh+2000} becomes an extremely tight power law $M_{\rm bar} \propto V^{\alpha'}$, with $\alpha'\sim 4$ and intrinsic scatter $\sim0.1$ dex, extending down to the lowest-mass dwarf galaxies \citep[e.g.,][]{McGaugh12,Lelli+2016,Lelli+2019}.
The Fall relation is also a power law in the form $j_\star \propto M_\star^\gamma$, with slope $\gamma\sim 0.6$ and a small intrinsic scatter of $\sim0.15$ dex, ranging from dwarf galaxies to massive discs \citep[$7\lesssim \log \mstar/\mo \lesssim 11$,][]{Romanowsky+2012,Posti+2018b,ManceraPina+2021,ManceraPina+2021b}.

In the standard $\Lambda$ cold dark matter ($\Lambda$CDM) paradigm, galaxies form within the gravitational potential of dark matter halos, whose assembly by hierarchical clustering is relatively well understood \citep[e.g.,][]{MovdBW10}. 
Because the $\Lambda$CDM perturbation spectrum is featureless and because gravity is a scale-free force, the global structure of halos is characterised by simple power-law scaling relations.
Some of these basic relations for dark matter halos are $M_{\rm h}\propto V_{\rm h}^3$, between halo mass $M_{\rm h}$ and virial velocity $V_{\rm h}$, and $j_{\rm h}\propto M_{\rm h}^{2/3}$, between halo specific angular momentum $j_{\rm h}$ and halo mass. While the $M_{\rm h}-V_{\rm h}$ relation follows from the definition of the virial radius, the $j_{\rm h}-M_{\rm h}$ relation is a consequence of the tidal torques exerted by surrounding halos and the resulting mass-independence of the spin parameter $\lambda$ \citep{Peebles+1969,Bullock+2001}.

The scaling laws of dark matter halos are analogous to those of disc galaxies, with also similar slopes, suggesting a co-evolution of dark and baryonic matter.
A convenient formalism that we can use to study galaxy scaling laws is to refer to the underlying relations for dark matter, and to focus on the ratios $f_X \equiv X_\star / X_{\rm halo}$ for a given quantity $X$, such as mass $M$, velocity $V$, or specific angular momentum $j$. 
With these ratios we can rewrite the Tully-Fisher and Fall relations referencing to the analogous relations for dark matter, as $M_\star \propto f_M (V_{\rm flat}/f_V)^3$ and $j_\star \propto f_j (M_\star/f_M)^{2/3}$ \citep[e.g.][]{Posti+2019}. 
If $f_M$, $f_j$, and $f_V$ were constants and did not scale with mass, then galaxies would be fully homologous to dark halos and the shape of their scaling relations would be identical. 
Conversely, any significant differences between the observed slopes of the Tully-Fisher and Fall relations and the dark halo values (3 and 2/3) would indicate that some combination of $f_M$, $f_j$, and $f_V$ must be mass dependent \citep[e.g.,][]{Navarro+00,DuttonvdBosch12,Ferrero+2017,Posti+2018a}.

In this framework, the remarkable similarity between galaxy and dark halo scaling laws is quite puzzling. 
This is because the relation between stellar mass and halo mass of the general population of galaxies, which can be expressed as the $f_M-M_\star$ or the $f_M-M_{\rm h}$ relation, is not a simple, featureless power-law relation. 
The current consensus is that $f_M$ increases up to a peak, at around $M_\star \sim 5\times 10^{10} \, \mo$ ($M_{\rm h} \sim 10^{12} \, \mo$), and then decreases with mass \citep[see e.g. the review by][]{Wechsler+2018}.
Such a highly non-linear relation should induce a prominent feature, similar to a break, on the scaling laws of disc galaxies at the mass scale where $f_M$ peaks. However, this is not observed, which implies that either the other ratios $f_j$ and $f_V$ are not constants \citep[][]{Cattaneo+14,Posti+2018a,Lapi+2018} or that the $f_M-M_\star$ relation of disc galaxies is actually a power law, contrary to estimates of the stellar-to-halo mass relation based on abundance matching \citep[][]{Posti+2019b,Posti+2019,PostiFall21}.

This issue has been studied by \cite{Lapi+2018}, using a compilation of stacked \ha\ rotation curves of 550 nearby disc galaxies. 
They find indications that the scaling laws have some deviations from power laws and that the ratios $f_M$, $f_V$, and $f_j$ are not constant, but they all increase with stellar mass up to a peak at $M_\star\sim 5\times 10^{10} \, \mo$, and then decrease with mass. 
Their \ha\ rotation curves, however, extend only up to $1-2$ optical radii of the galaxies, which is typically not enough to reach the flat and dark matter-dominated parts of the rotation curves \citep{vanAlbada+1985,Kent+1987,Kent+1988}, and are the result of stacking, which potentially biases the results as it attenuates individual differences \citep[e.g.][]{Noordermeer+2007}. 
\cite{Posti+2019} re-addressed this issue with a detailed analysis of a smaller sample of nearby discs with both extended atomic hydrogen (\hi) rotation curves and photometry at 3.6$\upmu$m.
Analysing each individual curve and surface brightness profile, they determined the disc scaling laws over a large mass range ($6.5\lesssim\log\,M_\star/M_\odot\lesssim 11.2$), finding no deviations from power laws, and showed that $f_j$ and $f_V$ are constant and close to unity across the entire mass range, while $f_M$ varies as a simple power law of \mstar. 
From these results they concluded that spirals are a self-similar population of objects, with the stellar-to-halo mass ratio $f_M$ scaling as a power law of $M_\star$ because of feedback from young stars \citep{Dekel+1986}.

A break in the scaling relations might still be present at higher masses than those explored by \cite{Posti+2019}, i.e.\ at $\log \mstar/\mo \gtrsim 11.2$. 
Finding such a feature would imply that the self-similarity of discs breaks at a larger mass scale than expected, suggesting that some physical process, such as inefficient cooling of proto-galactic gas, feedback from active galactic nuclei (AGN) or galaxy merging, are becoming important at that mass scale.
\cite{Ogle+2019} recently found indications for such a break in the Tully-Fisher and Fall relations of super-luminous spiral galaxies, a rare population of giant, star-forming discs with stellar masses $M_\star \simeq 2-7 \times 10^{11} \mo$ \citep{Ogle+2016,Ogle+2019a}. 
This study relied on long-slit \ha\ spectra for 23 massive spirals, from which they derived rotation velocities typically out to the optical radii. 
Since this result is potentially transformational for our understanding of disc galaxies, in this paper we set out to scrutinise in detail these findings through a more accurate characterisation of the kinematics and the dynamics of a larger sample of extremely massive spiral galaxies.
We develop a new technique to derive robust rotation curves through modelling of long-slit \ha-\nii\ emission-line observations. We use this technique to investigate the very high-mass ends of the Tully-Fisher and Fall relations and their connections to the relations at lower masses from previous works.

The remainder of this paper is structured as follows. 
Section~\ref{sec:data} introduces our galaxy sample and the long-slit data analysed in this work. 
Section~\ref{sec:profiles} and Section~\ref{sec:kinmod} describe the techniques used to derive stellar surface-density profiles and galaxy kinematics, respectively.
We present \ha\ rotation curves in Section~\ref{sec:rotcurs}, while in Section ~\ref{sec:scalrel} we build the Tully-Fisher and Fall scaling relations for massive spiral galaxies. We discuss our findings in Section~\ref{sec:discussion} and we conclude in Section~\ref{sec:conc}.
Throughout this paper, we use a flat $\Lambda$CDM cosmology with $\Omega_\mathrm{m,0} = 0.307$, 
$\Omega_\mathrm{m,0} = 0.693$ and $H_0 = 67.7$ km s$^{-1}$ Mpc$^{-1}$ \citep{PlanckCollaboration+2016}.
In this cosmology framework, 1 arcsec corresponds to 1.9 kpc and the lookback time is 1.3 Gyr at $z \simeq 0.1$.

\begin{table*}
\centering
\caption{Massive spiral galaxy sample analysed in this work. Columns: (1) Primary name from either 2MASS or SDSS surveys; (2) Alternative name used in this paper; (3) Redshift from kinematic modelling; (4) Stellar mass from $W1$ WISE aperture photometry (typical error is 0.2 dex); (5) Star formation rates from $W3$ WISE luminosity, with typical uncertainty of 30\%-40\%; (6) Inclination angle from $z$-band axis ratio (typical error is 5$\de$; (7) Position angle of the major axis (counterclockwise from North direction); (8) Velocity of the flat part of the rotation curve; (9) Warm-ionized gas velocity dispersion; (10) Total specific angular momentum. We also list galaxies excluded from our final sample, specifying the reason why we were not able to determine rotation curves for them.}
\label{tab:sample}
\begin{tabular}{llccccccccc}
\noalign{\vspace{1pt}}\hline\hline\noalign{\vspace{1pt}}
Name & Alt. name & $z$ & $\log \frac{M_\star}{\rm M_\odot}$ & SFR & $i$ & P.A. & $\vflat$ & $\vdisp$ & $\log \frac{\jstar}{\rm kpc \, km \, s^{-1}}$\\
 &  &  &  & $\moyr$ & deg & deg & km/s & km/s & \\
(1)  & (2) & (3) & (4) & (5) & (6) & (7) & (8) & (9) & (10)\\

\noalign{\vspace{1pt}}\hline\noalign{\vspace{1pt}}
2MASXJ00083819$-$0044064	&	-	&	0.03952	&	11.24	&	9	&	57	&	16	&	275	$\pm$	20	&	41	$\pm$	6	&	3.45	$\pm$	0.11	\\
2MASXJ01431323+0034405	&	-	&	0.17771	&	11.60	&	19	&	71	&	128	&	361	$\pm$	30	&	28	$\pm$	5	&	3.78	$\pm$	0.13	\\
2MASXJ02052519+0015129	&	-	&	0.17186	&	11.57	&	19	&	52	&	148	&	348	$\pm$	29	&	25	$\pm$	4	&	3.77	$\pm$	0.13	\\
2MASXJ02152347$-$0937283	&	-	&	0.21625	&	11.51	&	9	&	68	&	22	&	364	$\pm$	28	&	19	$\pm$	4	&	3.91	$\pm$	0.12	\\
2MASXJ02201242$-$0832228	&	-	&	0.11015	&	11.23	&	9	&	87	&	1	&	252	$\pm$	16	&	23	$\pm$	5	&	3.69	$\pm$	0.10	\\
2MASXJ02205358+0020002	&	-	&	0.13957	&	11.26	&	14	&	63	&	52	&	286	$\pm$	17	&	29	$\pm$	5	&	3.75	$\pm$	0.10	\\
2MASXJ09034365+0139334	&	-	&	0.10351	&	11.31	&	8	&	72	&	126	&	265	$\pm$	15	&	28	$\pm$	3	&	3.63	$\pm$	0.15	\\
2MASXJ09484652+0635428	&	-	&	0.08824	&	11.14	&	4	&	67	&	155	&	311	$\pm$	19	&	29	$\pm$	5	&	3.77	$\pm$	0.13	\\
SDSSJ095727.02+083501.7	&	OGC 0441 	&	0.25611	&	11.60	&	11	&	44	&	92	&	384	$\pm$	41	&	33	$\pm$	6	&	3.96	$\pm$	0.12	\\
2MASXJ09590646+0559293	&	-	&	0.09290	&	11.30	&	12	&	90	&	34	&	284	$\pm$	18	&	23	$\pm$	4	&	3.73	$\pm$	0.13	\\
2MASXJ10082752+0304597	&	-	&	0.16839	&	11.38	&	16	&	76	&	152	&	303	$\pm$	18	&	30	$\pm$	4	&	3.89	$\pm$	0.10	\\
2MASXJ10102901+0106588	&	-	&	0.10219	&	11.44	&	10	&	50	&	30	&	359	$\pm$	27	&	27	$\pm$	4	&	3.81	$\pm$	0.11	\\
2MASXJ10222648+0911396 	&	-	&	0.09143	&	11.42	&	17	&	71	&	75	&	307	$\pm$	21	&	25	$\pm$	7	&	3.70	$\pm$	0.10	\\
2MASXJ10304263+0418219	&	OGC 0926 	&	0.16105	&	11.66	&	34	&	50	&	64	&	325	$\pm$	28	&	28	$\pm$	5	&	3.70	$\pm$	0.11	\\
2MASXJ10494004+0026155	&	-	&	0.09103	&	11.14	&	6	&	83	&	102	&	270	$\pm$	17	&	27	$\pm$	5	&	3.61	$\pm$	0.10	\\
2MASXJ11052843+0736413	&	2MFGC 08638 	&	0.15244	&	11.59	&	24	&	89	&	80	&	458	$\pm$	21	&	26	$\pm$	6	&	4.16	$\pm$	0.09	\\
2MASXJ11193844+0047233	&	-	&	0.17419	&	11.44	&	15	&	60	&	129	&	413	$\pm$	24	&	25	$\pm$	3	&	3.81	$\pm$	0.10	\\
2MASXJ11232039+0018029 	&	-	&	0.14461	&	11.43	&	18	&	79	&	74	&	386	$\pm$	22	&	20	$\pm$	6	&	3.94	$\pm$	0.08	\\
2MASXJ11483552+0325268 	&	-	&	0.12007	&	11.42	&	13	&	76	&	29	&	293	$\pm$	18	&	25	$\pm$	5	&	3.84	$\pm$	0.10	\\
2MASXJ11495671+0457570	&	-	&	0.09303	&	11.25	&	13	&	80	&	14	&	260	$\pm$	21	&	28	$\pm$	4	&	3.61	$\pm$	0.10	\\
2MASXJ11535621+4923562	&	OGC 0586 	&	0.16673	&	11.64	&	38	&	66	&	128	&	338	$\pm$	27	&	20	$\pm$	5	&	3.89	$\pm$	0.16	\\
2MASXJ12003406+0510290	&	-	&	0.13060	&	11.29	&	10	&	80	&	123	&	273	$\pm$	19	&	29	$\pm$	5	&	3.72	$\pm$	0.10	\\
2MASXJ12422564+0056492 	&	-	&	0.07964	&	11.24	&	10	&	55	&	46	&	270	$\pm$	21	&	32	$\pm$	5	&	3.40	$\pm$	0.11	\\
2MASXJ12592630$-$0146580	&	-	&	0.08338	&	11.23	&	8	&	58	&	80	&	317	$\pm$	23	&	31	$\pm$	5	&	3.64	$\pm$	0.09	\\
2MASXJ13033075$-$0214004	&	2MFGC 10372 	&	0.08425	&	11.37	&	14	&	81	&	42	&	301	$\pm$	19	&	30	$\pm$	6	&	3.52	$\pm$	0.11	\\
2MASXJ13270366$-$0002093	&	-	&	0.14062	&	11.18	&	10	&	80	&	71	&	283	$\pm$	20	&	25	$\pm$	4	&	3.70	$\pm$	0.10	\\
2MASXJ13395130+0302174	&	-	&	0.11562	&	11.42	&	9	&	51	&	172	&	364	$\pm$	29	&	28	$\pm$	5	&	3.78	$\pm$	0.13	\\
2MASXJ13450793$-$0247470	&	-	&	0.07520	&	11.24	&	7	&	90	&	138	&	288	$\pm$	18	&	27	$\pm$	5	&	3.75	$\pm$	0.10	\\
SDSSJ143447.86+020228.6	&	OGC 1312 	&	0.27967	&	11.60	&	35	&	63	&	88	&	364	$\pm$	23	&	25	$\pm$	4	&	3.81	$\pm$	0.11	\\
2MASXJ15154614+0235564	&	2MFGC 12344 	&	0.14049	&	11.74	&	22	&	78	&	150	&	530	$\pm$	24	&	21	$\pm$	4	&	4.11	$\pm$	0.10	\\
2MASXJ15172226$-$0216475	&	-	&	0.11725	&	11.06	&	10	&	79	&	143	&	274	$\pm$	19	&	22	$\pm$	3	&	3.60	$\pm$	0.11	\\
2MASXJ15404057$-$0009331 	&	-	&	0.07849	&	11.39	&	13	&	66	&	44	&	310	$\pm$	21	&	22	$\pm$	6	&	3.65	$\pm$	0.09	\\
2MASXJ16184003+0034367 	&	-	&	0.16734	&	11.67	&	19	&	69	&	78	&	383	$\pm$	25	&	23	$\pm$	4	&	3.93	$\pm$	0.11	\\
2MASXJ16245634+0714325	&	-	&	0.15564	&	11.57	&	22	&	55	&	155	&	298	$\pm$	21	&	46	$\pm$	6	&	3.72	$\pm$	0.10	\\
2MASXJ20541957$-$0055204 	&	-	&	0.21014	&	11.41	&	24	&	63	&	2	&	316	$\pm$	22	&	19	$\pm$	4	&	3.77	$\pm$	0.10	\\
2MASXJ21175224$-$0057041	&	-	&	0.18022	&	11.38	&	19	&	78	&	174	&	303	$\pm$	22	&	20	$\pm$	4	&	3.82	$\pm$	0.10	\\
2MASXJ21362206+0056519 	&	-	&	0.10378	&	11.47	&	12	&	62	&	113	&	348	$\pm$	24	&	26	$\pm$	4	&	3.74	$\pm$	0.10	\\
2MASXJ21384311$-$0052162 	&	-	&	0.08306	&	11.20	&	3	&	58	&	78	&	291	$\pm$	23	&	28	$\pm$	5	&	3.56	$\pm$	0.11	\\
2MASXJ21431882$-$0820164 	&	-	&	0.06249	&	11.13	&	4	&	75	&	82	&	310	$\pm$	20	&	22	$\pm$	5	&	3.57	$\pm$	0.09	\\
2MASXJ21445295$-$0804037	&	-	&	0.12781	&	11.35	&	7	&	38	&	187	&	288	$\pm$	26	&	25	$\pm$	4	&	3.62	$\pm$	0.18	\\
2MASXJ22073122$-$0729223 	&	-	&	0.06332	&	11.20	&	10	&	71	&	84	&	223	$\pm$	19	&	33	$\pm$	6	&	3.45	$\pm$	0.09	\\
2MASXJ22520878+0015154	&	-	&	0.15176	&	11.25	&	12	&	80	&	45	&	292	$\pm$	19	&	24	$\pm$	5	&	3.62	$\pm$	0.10	\\
2MASXJ23130513$-$0033477 	&	-	&	0.11116	&	11.20	&	11	&	56	&	56	&	279	$\pm$	27	&	29	$\pm$	5	&	3.48	$\pm$	0.11	\\
\noalign{\vspace{0pt}}\hline\hline
\noalign{\vspace{7pt}}
Discarded galaxies \\
\noalign{\vspace{1pt}}\hline\noalign{\vspace{1pt}}
2MASXJ09394584+0845033$^\dagger$      & - & 0.13784 & 11.45 & 28 & 55 & 120 & \multicolumn{3}{l}{Highly asymmetric kinematics }\\
2MASXJ13180708+0502353      & - & 0.09626 & 11.26 & 5 & 87 & 39 & \multicolumn{3}{l}{Highly asymmetric kinematics}\\
2MASXJ13451949+0058117      & - & 0.16447 & 11.31 & 7 & 79 & 146 & \multicolumn{3}{l}{Low S/N (marginal detection)} \\
2MASXJ15592695+0842570      & - & 0.19962 & 11.52 & 14 & 46 & 59 & \multicolumn{3}{l}{Low S/N (marginal detection)} \\
2MASXJ16014061+2718161$^\dagger$      & OGC 1304 & 0.16466 & 11.63 & 23 & 60 & 116 & 
\multicolumn{3}{l}{Asymmetric: approaching side not well defined}\\
2MASXJ16262411+0841036      & - & 0.21553 & 11.43 & 9 & 72 & 28 & \multicolumn{3}{l}{Asymmetric: receding side not well defined}\\
2MASXJ16394598+4609058$^\dagger$     & OGC 0139 & 0.24714 & 11.74  & 45 & 76 & 115 & \multicolumn{3}{l}{Low S/N + anomalous emission} \\
2MASXJ21020534$-$0647558  & - & 0.12730  & 11.84  & 59 & 52 & 153 & \multicolumn{3}{l}{Low S/N + AGN contamination} \\
\noalign{\vspace{1pt}}\hline
\noalign{\vspace{2pt}}
\multicolumn{9}{l}{$^\dagger$ Galaxy in common with \citet{Ogle+2019}.}\\
\multicolumn{9}{l}{$^*$ Determination of \vflat\ may be affected by atmospheric absorption.}\\
\end{tabular}
\end{table*}

\begin{figure*}
	\includegraphics[width=0.99\textwidth]{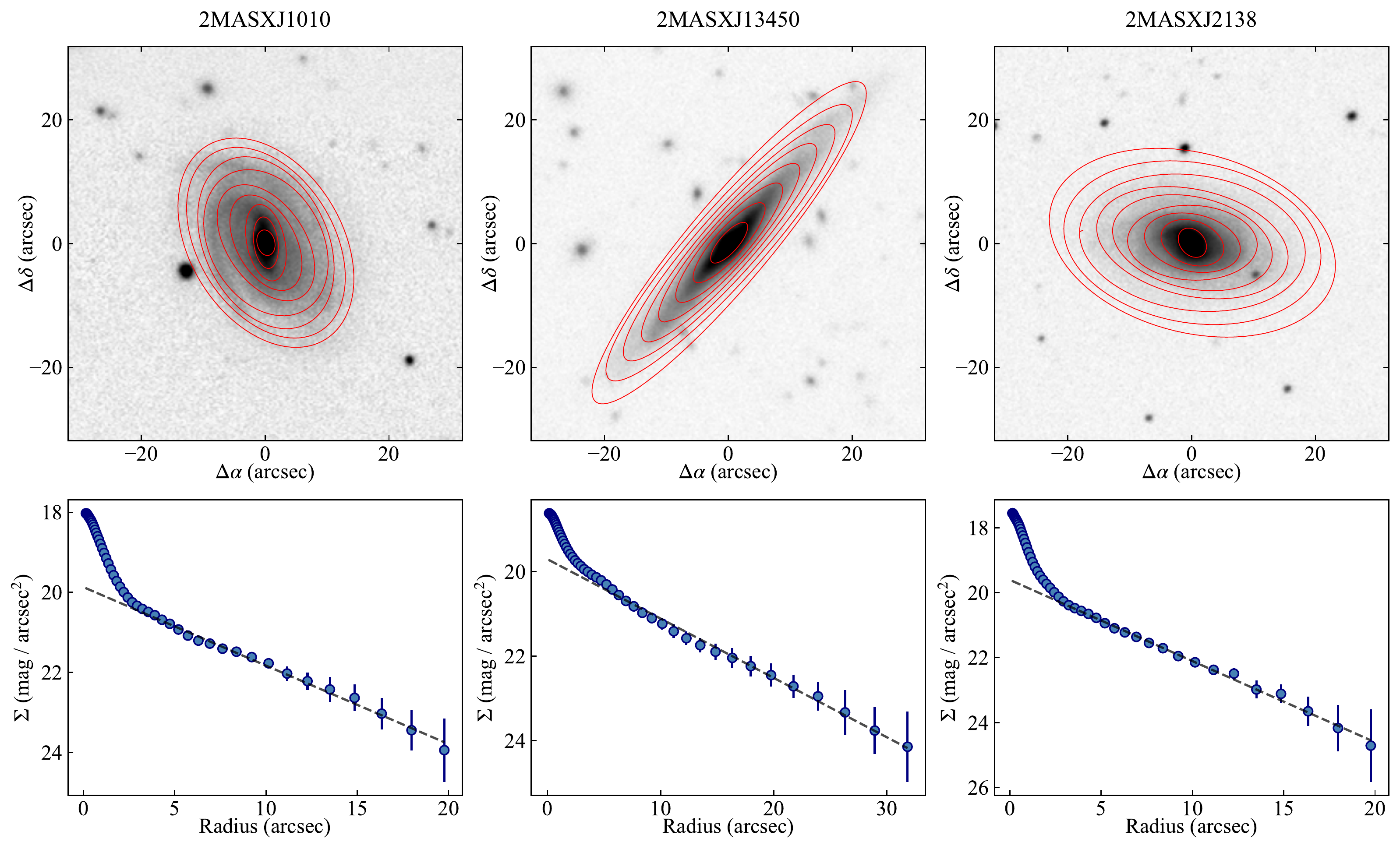}
    \caption{ Surface photometry for three massive spiral galaxies in our sample. \emph{Top panels}: $z$-band images from the Legacy survey \citep{Dey+2019} with overlayed some of the best-fit ellipses in red. \emph{Bottom panels}: average surface-brightness over the best-fit ellipses. The dashed gray line denotes an exponential function fit to the outer parts of the surface-brightness profile.}
    \label{fig:profiles}
\end{figure*}


\section{Galaxy sample and kinematic data}
\label{sec:data}

Our work is based on a sample of 51 massive star-forming galaxies for which we have collected long-slit spectroscopy observations.
Twenty-three galaxies are in common with \citet{Ogle+2019}.
The parent sample of massive spiral galaxies was selected from a combination of the most optically luminous Sloan Digital Sky Survey \citep[SDSS,][]{Eisenstein+2011} galaxies \citep[OGC catalog,][]{Ogle+2019a} and a lower-luminosity sample of infrared-selected galaxies \citep{Ogle+2019} from the 2-Micron
All-Sky Survey (2MASS) Extended Source Catalog \citep[2MASX,][]{Jarrett+2000,Skrutskie+2006} consisting of about 54 times as many galaxies.  
The SDSS sub-sample was selected from the $\sim$1600 most optically luminous galaxies of all types at redshift $z<0.3$, based on SDSS redshift and $r$-band photometry \cite[see][for details]{Ogle+2019a}.
The 2MASX sub-sample was selected from the $\sim$86000 galaxies of all types that have SDSS redshifts and $K_s$-band luminosity $L_{K_s}>2\times10^{11} \, L_\odot$.
The 2MASX sample was further restricted to the 19542 massive, equatorial galaxies within the SDSS footprint with declination $\delta_\mathrm{J2000} < 9\de$ that are observable by the Southern African Large Telescope \citep[SALT,][]{Buckley+2006}. 
To ensure reliable de-projected rotation velocities, we made a cut on galaxy ellipticity $\epsilon$, using bulge-disc decomposition from \citet{Simard+2011}, which yielded 4830 high-$\epsilon$ ellipticals, lenticulars, and spiral galaxies with inclination angle $i>40\de$. 
We visually identified 237 spiral galaxies in this subset by inspecting 3-color SDSS DR13 images using the Sky Server Chart service. 
Our final sample was selected from this final subset with preference for the largest actively star-forming spirals.
All 51 galaxies in our sample have SDSS images in $u$-$g$-$r$-$z$-$i$ filters, deeper $g$-$r$-$z$ images from the Dark Energy Spectroscopic Instrument (DESI) Legacy Surveys \citep[Legacy, hereinafter,][]{Dey+2019} and $J$-$H$-$K_\mathrm{s}$ band images from 2MASS.

To derive galaxy kinematics, we use long-slit observations of the \ha\ emission line at $6562.80 \ang$ and of the \nii\ emission-line doublet at $6548.05-6583.45 \ang$.
Long-slit data for 48 galaxies were obtained with the Robert Stobie Spectrograph \citep[RSS,][]{Burgh+2003} on SALT, using the 1800 line mm$^{-1}$ volume-phase holographic grating, which gives a resolving power of $4200-5300$ at the wavelengths corresponding to the redshifted \ha-\nii\ lines.
Three galaxies were observed with the Double Spectrograph (DBSP) on the Palomar Hale Telescope, using the 1200 line mm$^{-1}$ grating with a constant resolving power of 6000.
The width of the long-slit is 1$''$, while the typical seeing of observations is $1.5''-2''$.
We refer to \citet{Ogle+2019} for more details about the instrumental setup and data reduction.

Because only emission lines are required for our kinematic analysis, the galaxy continuum was removed from the long-slit observations: at each spatial position, a 1$^\mathrm{st}$-degree polynomial was fitted to the continuum emission in the spectral range free from the emission lines. 
The best-fit polynomials were then subtracted from the data, leaving only the emission from the \ha-\nii\ lines.
The \ha-\nii\ emission lines in these galaxies typically extends out to 2-4$R_\mathrm{d}$, where $R_\mathrm{d}$ is the scale radius of an exponential disc.
From our initial sample of 51 galaxies, we excluded 8 galaxies for which we could not derive robust rotation curves because of either strong asymmetries in their emission-line profiles or insufficient signal-to-noise ratio in the data.
This leaves us with a final sample of \galnum\ massive spiral galaxies, having redshift in the range $0.04\lesssim z\lesssim0.28$ and morphology spanning from early-type discs (e.g.\ Sa, Sab) to late-type discs (e.g., Sb, Sc).
The main properties of the sample are listed in \autoref{tab:sample}. 
Galaxies excluded from our kinematic analysis are also listed at the bottom of \autoref{tab:sample} and their \ha\ data can be found in \autoref{fig:discarded} of Appendix~\ref{appendixA}.
In the remainder of the paper, we will refer to individual galaxies in our sample by using either the alternative names listed in the second column of \autoref{tab:sample} or by cutting down the 2MASS name to the first four digits of the sky coordinates (e.g., 2MASXJ1022).


\section{Stellar masses and surface-density profiles}
\label{sec:profiles}

$W1$-band photometry at 3.4 $\upmu$m from the Wide-field Infrared Survey Explorer \citep[WISE,][]{Wright+2010} was used to calculate the total stellar content of our galaxies, applying the data pipeline and the methods described in \citet{Jarrett+2019} (see their Section 2.4).
Foreground stars and other contaminating sources were identified and removed from the images and the background was measured in an elliptical annulus and subtracted. 
Total flux was measured by fitting the radial surface brightness profile with a double Sérsic function and integrating the extrapolated profile out to 3$R_\mathrm{d}$, where $R_\mathrm{d}$ is the disc scale length. 
This is never more than 10\% more than the isophotal flux measured at the 1$\sigma$ outer isophote.
The $W1$-band flux was $K$-corrected using the 2MASS + WISE spectral energy distribution and used to compute the $W1$ luminosity $L_\mathrm{W1}$ relative to the Sun \citep{Cluver+2014}. 
Finally, the total mass in stars $\mstar$ was estimated assuming a constant mass-to-light ratio $\mstar/L_\mathrm{W1}=0.6$ with a typical uncertainty of $\sim40\%$ \citep{Meidt+2014,Norris+2014,Rock+2015}.
Errors on the final stellar masses are of the order 0.2 dex and are dominated by the uncertainty on the mass-to-light ratio. 
Dust obscuration is expected to be negligible or small at 3.4 $\upmu$m.
The fourth column of \autoref{tab:sample} lists our derived stellar masses $\mstar$. 
All galaxies in our sample have stellar masses that exceed $10^{11} \, \mo$, with the most massive one reaching a $\mstar\simeq6\times10^{11} \, \mo$.
Star formation rates (SFRs) for our galaxies were estimated from the WISE $W3$-band at 12 $\upmu$m, following the prescriptions of \citet{Cluver+2017} (see their Equation 4). Our galaxies have dust-obscured SFRs ranging from a few $\moyr$ up to a few tens $\moyr$ (column 5 in \autoref{tab:sample}).

We use $z$-band images from the Legacy surveys to derive stellar surface-density profiles $\sigmastar (R)$, under the assumption that light traces stellar mass. 
Surface photometry analysis is performed using the \textsc{photoutils} package \citep{larry_bradley_2020_4044744} implemented within the \textsc{astropy} Python library \citep{astropy:2013,astropy:2018}. 
A median sky value is determined after masking contaminating sources and subtracted from each image. 
We then measure azimuthally-averaged surface-brightness profiles by fitting a set of elliptical isophotes to galaxy images, using the iterative method described in \citet{Jedrzejewski+1987}. 
Each ellipse is defined by four parameters: the galaxy centre $(x_0,y_0)$, the position angle $\phi$ of the ellipse's major axis with respect to the North direction and the ellipticity $\epsilon\equiv1-b/a$, where $b/a$ is the ratio of the minor axis $b$ to the major axis $a$. 
We provide an initial guess ellipse and we let all parameters free during the ellipse fitting. 
During the procedure, pixels deviating more than 3$\sigma$ from the average value within an ellipse, which may represent unresolved background/foreground objects, were automatically masked. 
Total errors for each ellipse were calculated as $\sigma^2_\Sigma(R)=\sigma^2_\mathrm{rms}(R) + \sigma^2_\mathrm{sky}$, where $\sigma_\mathrm{rms}(R)$ is the root-mean-square (rms) variation around each ellipse and $\sigma_\mathrm{sky}$ is the error on the sky determination. 
Typically, the first term dominates the error budget in the inner regions of a galaxy, while the latter dominates in the outermost regions. 
Best-fit position angles of the outermost isophotes generally agree with the long-slit position angles within a few degrees.
\autoref{fig:profiles} illustrates examples of surface-density photometry for three galaxies in our sample. 
Bottom panels show their stellar surface-brightness profiles derived from the $z$-band images along the best-fit ellipses (red lines) displayed in the top panels.


\section{Kinematic modelling}
\label{sec:kinmod}
We exploited a hybrid 3D-1D approach to derive accurate rotation curves from our long-slit spectroscopic data. 
Our procedure starts with building three-dimensional emission-line models through the kinematic code \bba\ \citep{DiTeodoro&Fraternali15}.
\bba\ uses a 3D tilted-ring model to simulate 3D emission-lines spectral datacubes (2 spatial and 1 spectral dimension) for a galaxy with given geometrical and kinematic parameters. 
The geometry is characterized through five parameters: the central positions of the galaxy $(x_0,y_0)$, the inclination angle $i$ with respect to the line of sight (90$^\circ$ for an edge-on disc), the position angle $\phi$ of the disc's major axis with respect to the North direction, and the scale height $Z_0$ of the disc. 
The kinematics is defined by three main parameters: the redshift $z$, the rotation velocity $\vrot$, and the intrinsic gas velocity dispersion $\vdisp$.
A 3D modelling allows to account easily for the instrumental biases introduced by the Point Spread Function (PSF) of the telescope, which can be important for galaxies resolved with just a few resolution elements \citep[beam smearing, e.g.][]{Begeman+1987}, and by the spectral resolution of the spectrograph. 
In addition, many of our galaxies have $i>70^\circ$ and a 3D model allows us to deal with highly-inclined systems, overcoming the well-known issues of deriving rotation curve in nearly edge-on galaxies \citep[e.g.,][]{Fraternali+11}.

Throughout the fitting procedure, we constructed 3D models using ring widths of 1.5$''$ and fixing the galaxy geometry. 
The galaxy centre was set to the position corresponding to the peak flux of the continuum emission in the long-slit data.
The position angle was arbitrarily fixed to $\phi=90^\circ$ and the disc thickness to $Z_0=500$ pc. 
We stress that the discs of these galaxies seem to be extremely thin, as demonstrated by our edge-on systems' optical images, implying that the effect of the disc scale height on the derived rotation curve is most likely negligible.
The inclination angle was calculated from the axis ratios from the isophotal fit of the optical images (Section~\ref{sec:profiles}), i.e.\ $\cos^2 i= [({b/a})^2 - q_0^2]/(1-q_0^2)$, where $a$ and $b$ are the major and minor semi-axes, respectively, and we assume $q_0=0.16$ for these thin discs \citep[e.g.,][]{Fouque+1990}.
To avoid issues with the possible presence of a spheroidal bulge in the inner regions of our galaxies, we used a $b/a$ averaged over the outermost isophotes only.
Keeping fixed the galaxy geometry to the above parameters, we fitted for the galaxy redshift $z$, the rotation velocity at each radius $\vrot(R)$ and a constant value for the gas velocity dispersion $\vdisp$.

\begin{figure*}
	\includegraphics[width=0.98\textwidth]{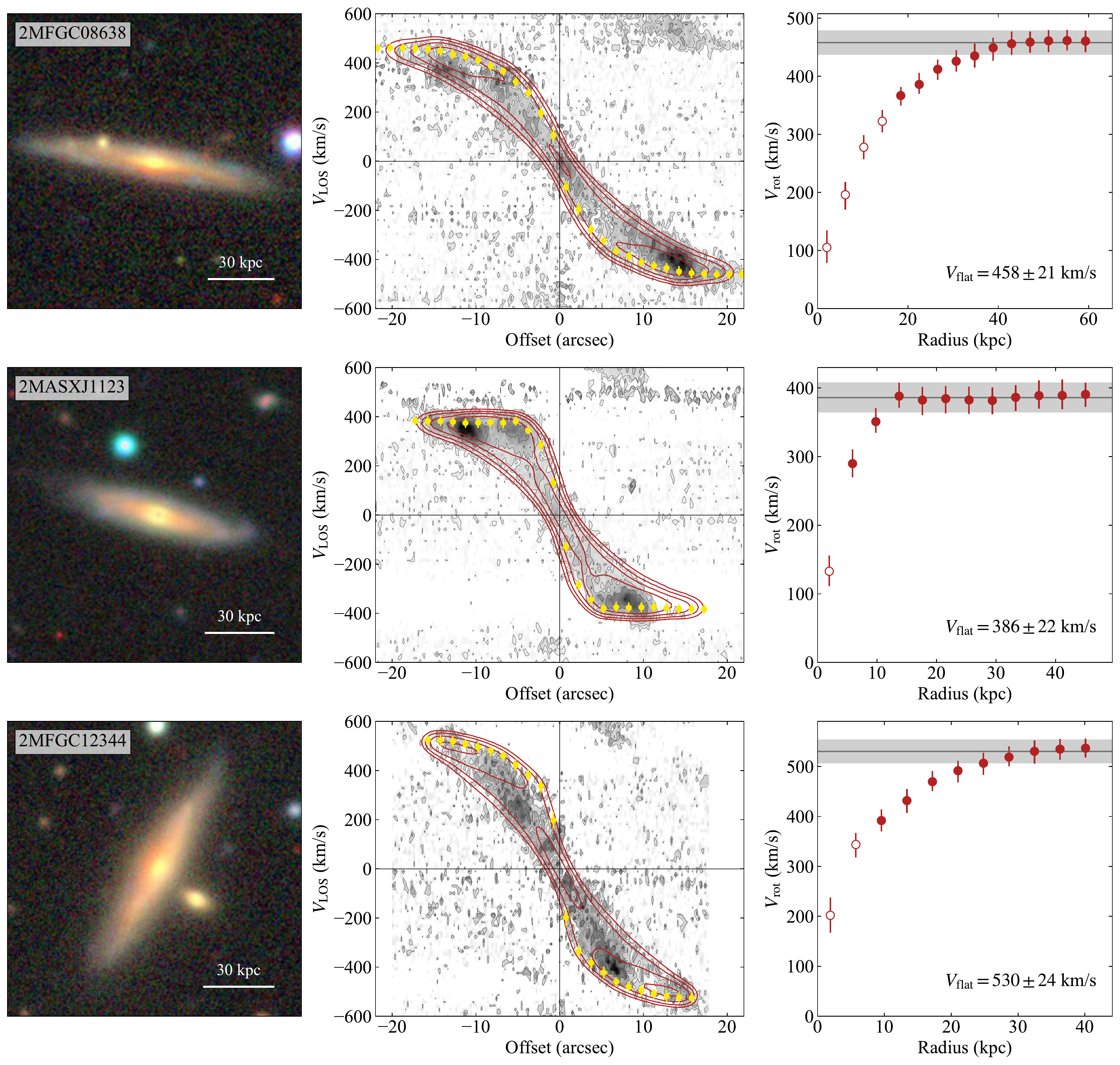}
    \caption{Rotation curves for our three fastest rotation galaxies. \emph{Left:} RGB images in the $g-r-z$ bands from the Legacy surveys \citep{Dey+2019}.
    \emph{Middle}: continuum-subtracted \ha\ long-slit data from SALT, shown in grey scale and contours, and our best-fit kinematic models (red contours). 
    Some of the \nii-doublet emission is also visible at the top/bottom edges of the plots. Contours are drawn at $2^n \times 2.5\sigma_\mathrm{rms}$, where $\sigma_\mathrm{rms}$ is the rms noise of the data and $n=0,1,2,3,4$. Yellow dots denote the derived rotation velocities projected along the line of sight (i.e.\ $\vrot\sin i$).
    \emph{Right}: Best-fit rotation curves (red circles). Errors in the rotation curves include systematic uncertainties on the inclination angle and long-slit positioning. White empty dots highlight radii where the rotation velocity can not be robustly determined because of possible projection/absorption effects. 
    The grey horizontal line and band represent the velocity of the flat part of the rotation curve and its error, respectively.}
    \label{fig:maps}
\end{figure*}

Model datacubes were simulated using a spatial pixel size of $0.254''$ and a spectral channel width of $0.3-0.4 \, \ang$, corresponding to the spatial and spectral pixels of the long-slit data, respectively.
An instrumental spectral broadening of $\sigma_\mathrm{instr}\simeq1\,\ang$ was added in quadrature to the gas intrinsic velocity dispersion \vdisp\ to take into account the finite resolving power of spectrographs.
Three-dimensional models were convolved with a circular 2D Gaussian kernel with Full Width at Half Maximum (FWHM) of 1.5$''$, representing the typical seeing of long-slit observations \citep{Ogle+2019}. 
From these 3D model datacubes, we extracted a $1''$-wide position-wavelength slice along the kinematic major axis, representing a simulated long-slit observation, and we fitted the model parameters $\mathbf{p}= \{z,\vrot(R),\vdisp\}$ by minimizing the absolute residuals between the simulated and the real data:

\begin{equation}
\mathcal{F}(\mathbf{p}) = \frac{1}{N}\sum_{x,\lambda} \mid f_\mathrm{data}(x,\lambda) - f_\mathrm{model}(x,\lambda,\mathbf{p}) \mid 
\end{equation}

\noindent where $f_\mathrm{data}(x,\lambda)$ and $f_\mathrm{model}(x,\lambda)$ are the observed and predicted flux densities at a given position $x$ and wavelength $\lambda$, respectively, and the summation extends over all the $N$ pixels with signal-to-noise ratio $\mathrm{S/N}>2.5$ in the continuum-subtracted slit data. 
Absolute residuals give more weight to regions where the emission is faint and diffuse with respect to the square of residuals \citep{DiTeodoro&Fraternali15}.
The optimization of the objective function $\mathcal{F}(\mathbf{p})$ made use of a Multilevel Coordinate Search (MCS) algorithm \citep{Huyer&Neumaier98}.
The above procedure was applied preferentially to the \ha\ emission line only, which is usually stronger than the \nii-doublet. 
For two galaxies (OGC0586 and 2MASXJ2138), for which part of the \ha\ line was affected by the subtraction of atmospheric sky lines, we fitted our kinetic model to three emission lines simultaneously \cite[e.g.,][]{DiTeodoro+2018}.

Uncertainties on the derived parameters, in particular on the rotation curve \vrot($R$), are dominated by 1) the relatively-low S/N of the data, 2) systematic errors on the assumed inclination angle $i$ and 3) possible misalignment of the long-slit with respect to a galaxy's major axis (i.e.\ an error on the position angle $\phi$).
To take into account these three important effects, we estimated parameter uncertainties via bootstrapping.
We produced 10k realizations of our best-fit parameters: in each realization, a $1\sigma_\mathrm{RMS}$ Gaussian noise was added to the long-slit data, while $i$ and $\phi$ values were drawn from normal distributions with mean given by our fiducial values and standard deviation of $5^\circ$. 
We checked that bootstrap distributions of our free parameters were nearly Gaussian after 10k iterations and we assumed the values at $15.87^\mathrm{th}$ and the $84.13^\mathrm{th}$ percentiles as 1$\sigma$-deviation errors.


\section{Rotation curves and gas velocity dispersion}
\label{sec:rotcurs}

\autoref{fig:maps} illustrates our best-fit models and the corresponding \ha\ rotation curves for three fast rotators in our sample. 
Models for all \galnum\ galaxies can be found in \autoref{fig:appfig} of Appendix~\ref{appendixA}. 
We show composite $g,r,z$ images from the Legacy surveys (\emph{left} panels), the \ha\ emission along the major axes from our long-slit observations (\emph{middle} panels) and our best-fit rotation curves (\emph{right} panels). 
Grey contours in middle panels denote the data, red contours our best-fit models, with yellow points representing the rotation velocities projected along the line of sight. 
Empty points in the rightmost panels highlight regions where the derived rotation velocities is less reliable due to possible extinction and projection effects that are not taken into account during the kinematic modelling. 
Unlike the atomic hydrogen line at 21 cm (\hi), the \ha\ line is not optically thin and extinction may occur, especially in the innermost regions of highly-inclined systems.
For example, a hole in the ionized gas distribution or an extincted region with a high rotation velocity in the inner parts of a nearly edge-on galaxy may be hardly visible in our \ha\ data. 
To better understand these effects, we build a set of kinematic models starting from our best-fit rotation curve and by progressively removing emission from the inner rings of each galaxy until the model starts to deviate significantly from the data. 
This procedure allows us to identify the minimum radius where the rotation curve is reliable. 
Finally, for galaxies where the line emission is considerably non-axisymmetric with respect to the galaxy centre, we preferentially fit only the side that is either more regular or more extended in radius. An example of a galaxy with asymmetric emission is 2MASXJ1123, visible in the second row of \autoref{fig:maps}.

All galaxies in our sample have rotation curves that reach a flat part in the outer regions of the optical disc. 
The shape of rotation curves of spiral galaxies have been observed to have a clear dependence on galaxy luminosity \citep[e.g.,][]{Persic+1996}. In particular, massive spiral galaxies typically have rotation curves that sharply rise in the inner regions and then quickly approach an asymptotic value. 
Some earlier-type massive disc galaxies are known to exhibit slowly declining rotation curves that flatten out at large radii \citep[e.g.,][]{Kent+1988,Noordermeer+2007}. 
As expected, most galaxies in our sample show steeply rising rotation curves that flatten well within 2-3$R_\mathrm{e}$, where $R_\mathrm{e}$ is the effective radius. 
Notable exceptions are our two fastest rotating discs, 2MFGC08638 and 2MFGC12344 (see first and third rows in \autoref{fig:maps}).
Although the inner rotation velocities of these galaxies could be very uncertain because of their high inclination angles, the rotation curves seem to rise slowly and reach a flat region only in the outermost regions of the stellar disc.
This evidence might indicate that the dynamics of these galaxies is dominated by the dark matter halo even in the inner regions.
These two galaxies also reach extremely large rotation velocities, with 2MFGC12344 arguably setting the record for the most rapidly rotating spiral galaxy currently known \citep[see also UGC12591,][]{Giovanelli+1986}.
Some galaxies in our sample have rotation curves that peak very near their centres (e.g., 2MFGC10372, 2MASXJ1200), highlighting the presence of massive bulges that are also visible in the optical images. 
Although a few galaxies show a slight decline in some part of the optical rotation curve, none seems to have a rotation velocity that is still declining significantly at the outermost measured radius, with the possible exceptions of 2MASXJ0205 and 2MASXJ0215.

\begin{figure*}
	\includegraphics[width=0.98\textwidth]{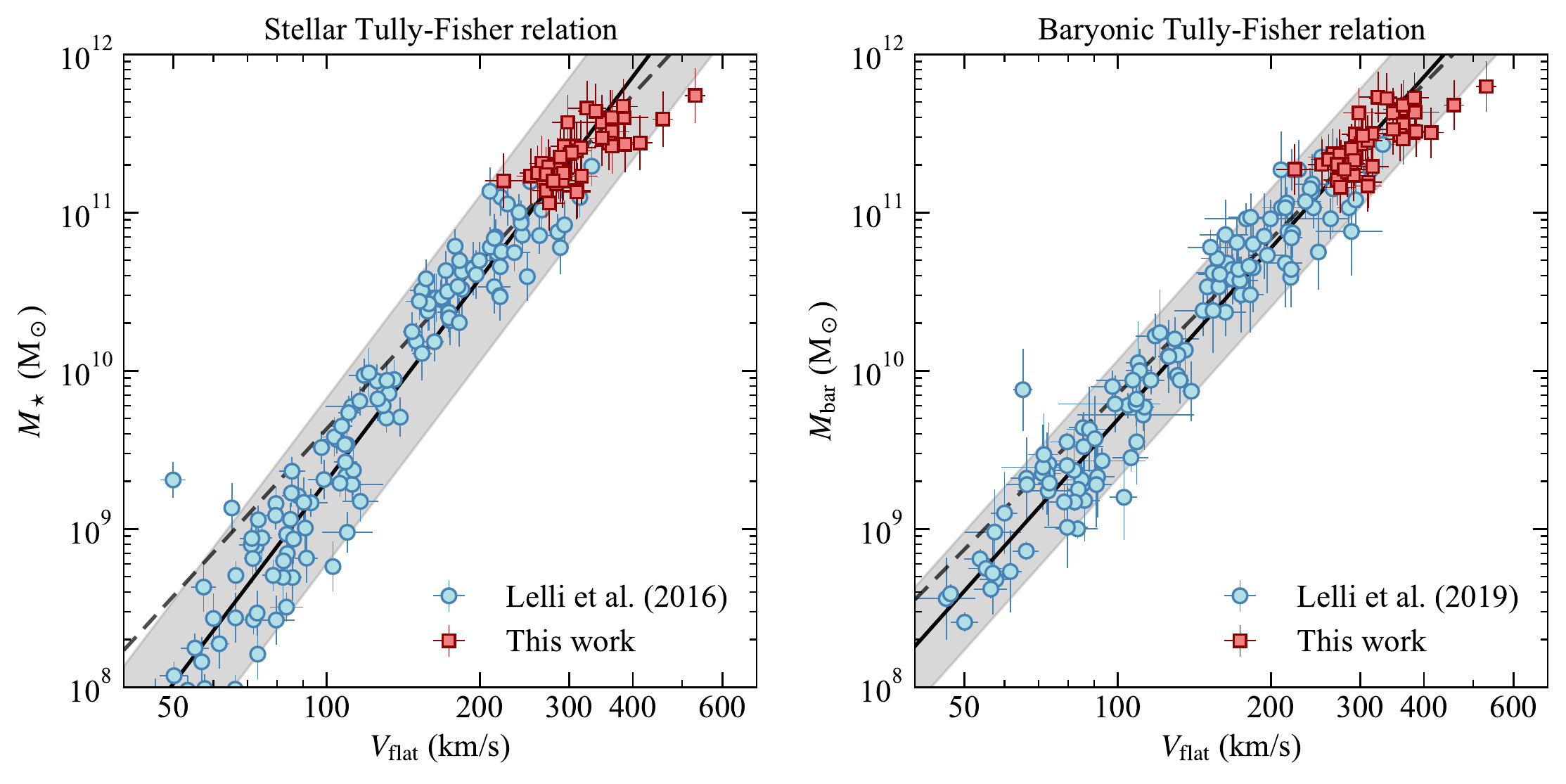}
    \caption{Tully-Fisher relations for massive spiral galaxies. The stellar Tully-Fisher relation (\emph{left}) and the baryonic Tully-Fisher relation (\emph{right}) are shown. 
    We plot our sample of massive spirals (red squares) and galaxies from the SPARC catalog (cyan circles) \citep{Lelli+2016b}.
    Solid and dashed black lines are linear fits to all galaxies and to galaxies with stellar/baryonic masses larger than $3\times10^{10} \, \mo$, respectively. 
    Best-fit parameters for the relation in the form $\log M = \alpha\log \vflat + \beta$ are $(\alpha,\beta)=(4.25,0.80)$ (full sample) and $(\alpha,\beta)=(3.51,2.61)$ (high mass) for the stellar Tully-Fisher relation, and $(\alpha,\beta)=(3.60, 2.49)$ (full sample) and $(\alpha,\beta)=(3.26,3.33)$ (high mass) for the baryonic Tully-Fisher relation. 
    Grey bands denote the 1$\sigma$ orthogonal scatter of data around the best-fit relations for all galaxies, i.e.\ 0.15 dex for the stellar Tully-Fisher relation and 0.1 dex for the baryonic Tully-Fisher relation.}
    \label{fig:TF}
\end{figure*}

Together with the rotation velocity, our kinematic modelling returns the velocity dispersion of the warm ionized gas that originate the \ha-\nii\ emission lines. 
Our best-fit value is corrected for observational biases, in particular for the PSF and for the instrumental broadening, and therefore represents the intrinsic velocity dispersion of gas, accounting for thermal and turbulent broadening. 
We find that the 1D velocity dispersion in our sample ranges between 22 km/s to 33 km/s (see \autoref{tab:sample}), values fully consistent with previous measurements of gas velocity dispersion of the warm ionized medium in less massive spiral galaxies \citep[e.g.,][]{Andersen+2006,Martinsson+2013}. 
This implies that the kinematics of these systems is strongly dominated by rotational motions, with $\vrot/\vdisp\gg 10$ for all galaxies.
Overall, from the shape of the rotation curves and from the gas velocity dispersions, we can conclude that these spiral galaxies seem akin to less massive discs from a kinematical point of view.


\section{Scaling relations}
\label{sec:scalrel}

\subsection{Tully-Fisher relations}

The Tully-Fisher relation comes in different flavors, depending on the way mass, luminosity or rotation velocity are measured. 
Historically, luminous matter has been expressed through either galaxy luminosity in different bands \citep[e.g.,][]{Karachentsev+2002}, stellar mass \citep[e.g.,][]{Torres-Flores+2011}, or baryonic mass \citep[stars + cold gas, e.g.][]{McGaugh2005}.
Widely-used velocity definitions \citep[see e.g.,][for a detailed discussion]{Lelli+2019} include the width of the global \hi\ profile $W_\hi$, the circular velocity $V_\mathrm{2.2}$ at 2.2$R_\mathrm{d}$, the maximum rotation velocity $V_\mathrm{max}$ and the mean velocity along the flat part of the rotation curve $V_\mathrm{flat}$.
Several studies have shown that the latter velocity definition, $\vflat$, minimizes the scatter of the relation \citep{Verheijen+2001,Lelli+2019}.

We use the stellar masses derived in Section~\ref{sec:profiles} and the flat part of the rotation curves modelled in Section~\ref{sec:kinmod} to build the stellar Tully-Fisher relation $\mstar - \vflat$ for our massive spiral galaxies. 
Following \citet{Lelli+2016}, we calculate $\vflat$ using a simple iterative procedure that quantifies the level of flatness of a curve.
This algorithm starts from the last points of a rotation curve and proceeds inwards, evaluating the variation in velocity with respect to the average flat velocity $V_\mathrm{a}$ until a deviation larger than 5\% is found \citep[for details, see Section 2.2 of][]{Lelli+2016}. 
All galaxies in our sample happen to have rotation curves that are flat within 5\% over at least three velocity points. 
Under the assumption that rotation curves remain flat at larger radii, our measured $\vflat$ corresponds to the asymptotic rotation velocity.
The uncertainty on $\vflat$ is calculated as: 

\begin{equation}
\label{eq:errvflat}
    \delta_{V_\mathrm{flat}} = \left( \frac{1}{N}\sum_i^N \delta^2_{V_i} + \delta^2_{V_\mathrm{a}} \right)^{\frac{1}{2}}
\end{equation}

\noindent where $\delta_{V_i}$ is the error associated with each of the $N$ velocity points along the flat part of the rotation curve and $\delta_{V_\mathrm{a}}$ is the dispersion around the average flat velocity value, which quantifies the degree of flatness of the curve. 
We note that the errors on inclination and position angles are already included in $\delta_{V_i}$.
Derived $\vflat$ are listed in \autoref{tab:sample} (column 8).
The gray lines and shaded regions in the rotation curves (rightmost panels) of \autoref{fig:maps} and of \autoref{fig:appfig} denote the resulting $\vflat$ and associated error, respectively. 

The left panel of \autoref{fig:TF} shows our determination of the stellar Tully-Fisher relation. 
Red squares at the high-mass end of the relation highlight our galaxy sample. 
As a comparison, we plot with cyan circles the relation from the Spitzer Photometry \& Accurate Rotation Curves catalog \citep[SPARC,][]{Lelli+2016b}, which is the largest collection of galaxies with high-quality photometry at 3.6$\upmu$m and \hi\ rotation curves currently available. 
We fit a linear function in the form $\log \mstar = \alpha\log \vflat + \beta$ to the entire sample and to galaxies with $\log (\mstar / \mo) > 10.5$ only, using an orthogonal distance regression technique to take into account errors in both variables.
This allows us to quantify whether the relation changes slope at the high-mass end.
The solid line in \autoref{fig:TF} denotes the the best-fit linear relation to the full sample, $\alpha=4.25\pm0.19$ and $\beta=0.80\pm0.23$, with an orthogonal scatter of $\sigma_\bot \simeq 0.15$ dex (grey band). 
We note that the grey band in \autoref{fig:TF} is wider than the errors of individual galaxies because the term $\delta_{V_\mathrm{a}}$ (flatness deviation) dominates in Eq.~\ref{eq:errvflat}.
The dashed line is the best-fit to the high-mass sample only, $\alpha=3.51\pm0.22$ and $\beta=2.61\pm0.21$. 
These best-fit relations are consistent with each other and are in good agreement with recent determinations of the stellar Tully-Fisher relation \citep[e.g.,][]{Reyes+2011,Bekeraite+2016,Bloom+2017}.
From this we conclude that the relation seems to continue unbroken from dwarfs to super spirals.
The two marginal outliers visible in the left panel of \autoref{fig:TF} in our massive spiral sample are the two fastest rotators, 2MFGC08638 and 2MFGC12344. 

The baryonic Tully-Fisher relation is considered the straightest and tightest of all Tully-Fisher relations \citep{McGaugh+2000,McGaugh2005}. 
Unfortunately, there are no measurements of the cold gas content of the galaxies in our sample, and the strong radio-frequency interference at the redshifted \hi\ frequencies prevents such measurements.
We can however have a rough estimate of gas masses in these galaxies via the Kennicutt-Schmidt relation \citep{Schmidt+1959,Kennicutt+1998}. 
In particular, we use the global star formation law recently calibrated by \citet{delosReyes+2019} in local spiral galaxies:

\begin{equation}
    \log \left(\frac{\Sigma_\mathrm{SFR}}{\mathrm{\mo \, yr^{-1} \, kpc^{-2}}}\right) = 1.41\log \left(\frac{\Sigma_\mathrm{gas}}{ \mathrm{\mo \, pc^2}}\right) -3.84
\label{eq:ks}
\end{equation}

\noindent where $\Sigma_\mathrm{SFR}$ is the star formation rate density and $\Sigma_\mathrm{gas}$ is the cold gas surface density (atomic + molecular gas). 
We assumed a typical scatter of 0.3 dex for this relation. 
$\Sigma_\mathrm{SFR}$ was calculated by normalizing the SFR by the deprojected star-forming area $\pi R_{25}^2$, where $R_{25}$ is the radius at the 25 mag arcsec$^{-2}$ isophote in $z$-band. 
We then obtained the gas surface density $\Sigma_\mathrm{gas}$ via Eq.~\ref{eq:ks} and converted to gas mass $M_\mathrm{gas}$ by multiplying $\Sigma_\mathrm{gas}$ by the same area.
This results in typical gas masses $\log (M_\mathrm{gas}/ \mo) \simeq 10.0-10.9$ and gas fractions $M_\mathrm{gas}/\mstar \simeq 0.08-0.20$. 
We note that these cold gas masses calculated via the Kennicutt-Schmidt law are also in good agreement with the expectations from the empirical relation between stellar mass and \hi\ mass in gas-rich spiral galaxies \citep[e.g.][]{Parkash+2018,Naluminsa+2021}.
Finally, we calculate the total baryonic mass as  $M_\mathrm{bar} = \mstar\ + M_\mathrm{gas}$.

The right panel of \autoref{fig:TF} shows the baryonic Tully-Fisher relation for our galaxies (red squares) compared to the SPARC sample by \citet{Lelli+2019}.
As before, the black solid line is a best-fit to the entire mass range, $\alpha=3.60\pm0.17$ and $\beta=2.49\pm0.20$, which is virtually indistinguishable from previous determinations from SPARC \citep{Lelli+2016,Lelli+2019}. 
The black dashed lines is a best-fit to galaxies with $\log (M_\mathrm{bar} / \mo) \geq 10.5$ only, $\alpha=3.26\pm0.14$ and $\beta=3.33\pm0.19$.
Our galaxies are perfectly compatible with the relation determined by lower mass galaxies, similar to what we find for the stellar Tully-Fisher relation (\autoref{fig:TF}, left panel). The only two galaxies that can be considered outliers of the baryonic Tully-Fisher relation are, again, the two fastest rotating galaxies in our sample.
Although our gas mass estimates are rather crude, we are confident on our estimates of the total baryonic masses since stars always dominate the baryonic budget in giant spiral galaxies. Even if the gas masses were a factor of 2 larger/smaller than what we assume, our sample of galaxies would still be consistent with the scaling relations determined at lower masses.

\subsection{Fall relation}

For an axisymmetric disc rotating on cylinders about its symmetry axis, the specific stellar angular momentum within a radius $R$ from the centre can be written as: 

\begin{equation}
\label{eq:jstar}
    \jstar(<R) = \frac{J_\star(<R)}{\mstar(<R)} = \frac{\int_0^R \, \sigmastar(R') \, {R'}^2 \, V_\mathrm{\star,rot}(R') \,  dR' }{\int_0^R \, \sigmastar(R') \, R' \, dR' }
\end{equation}

\noindent where $\sigmastar$ and $V_\mathrm{\star,rot}$ are the mass surface density and the azimuthal velocity of the stellar component, respectively. Equation~\ref{eq:jstar} is used to compute the specific angular momentum as a function of radius for our galaxies. 
We use the stellar surface density profiles derived in Section~\ref{sec:profiles} and we assume $V_\mathrm{\star,rot}=V_\mathrm{\ha,rot}$. 
We expect any correction to the circular velocity for pressure support to be negligible for these highly-flattened disks, i.e. $\sigma_\star / V_\mathrm{\star, rot} \sim z_\mathrm{d}/R_\mathrm{d} \ll 1$, where $z_\mathrm{d}$ and $R_\mathrm{d}$ are the disk scale-height and scale-length, respectively \citep[see also][]{ManceraPina+2021}.

\begin{figure}
	\includegraphics[width=0.47\textwidth]{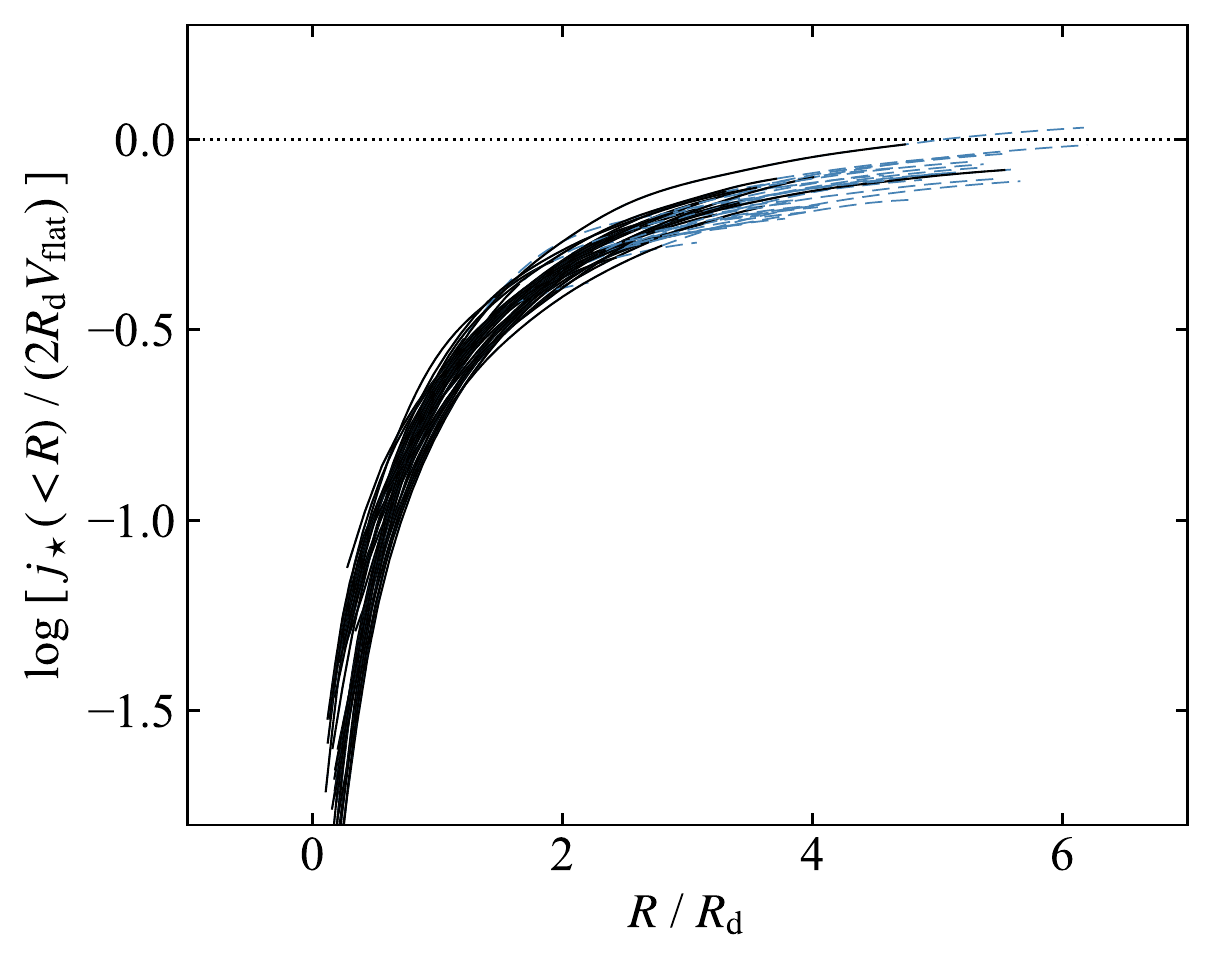}
    \caption{Stellar specific angular momentum profiles for the \galnum\ massive spiral galaxies in our sample. Black solid lines show the inner regions where both \ha\ rotation velocity and stellar surface-brightness are measured. These profiles are either fully converged or converging. Blue dashed lines show the extrapolation for a flat rotation curve and an exponentially declining surface-density profile. The radius $R$ is normalised to the disc scale length $R_\mathrm{d}$ in $z$-band and $\jstar$ to the specific angular momentum of a thin exponential disc with the same $R_\mathrm{d}$ and with a constant rotation curve (dotted line).}
   \label{fig:jprof}
\end{figure}

\autoref{fig:jprof} shows cumulative $\jstar$ profiles for our \galnum\ galaxies calculated through Equation~\ref{eq:jstar}.
Black solid lines denote $\jstar(<R)$ within the last point of the measured rotation curve. 
Most profiles are converging within the optical disc but are not fully converged to the asymptotic value representing the total specific angular momentum of the galaxy.
Therefore, we calculate $\jstar$ at larger radii by extrapolating the rotation curves and surface-brightness profiles beyond the last measured data point. 
Since all of our galaxies have reached the flat part of the rotation curve, as discussed in Section~\ref{sec:rotcurs}, we assume that the velocity remains constant at larger radii. 
For the surface brightness, we fit an exponential function to the outer regions of the optical disc (dashed lines in \autoref{fig:profiles}) and we assume that the exponential decline continues beyond the last measured point. 
Our extrapolation stops at the radius where the converging criterion used in \citet{Posti+2018b} is met, i.e.\ $\Delta \jstar / \jstar < 0.1$ and $\Delta\log\jstar/\Delta \log R < 0.5$, where $\Delta$ is measured over the two outermost radii. 
Blue-dashed lines in \autoref{fig:jprof} denote the extrapolated part of the converged $\jstar$ profiles.
Errors on the total specific angular momentum were calculated by bootstrapping 10k realization of each \jstar\ profile.
To take into account a possible decline or increase of the rotation curve beyond the optical disc, we draw random extrapolations of the rotation curve by allowing a conservative maximum variation of $20\%$ from $\vflat$. 
Errors on the surface brightness were taken into account by drawing random extrapolations around the best fit exponential function.
The $15.87^\mathrm{th}$ and the $84.13^\mathrm{th}$ percentiles of the resulting specific angular momentum distribution for each galaxy are  taken as 1$\sigma$ error. 
We find that typical uncertainties on $\jstar$ are of the order of 0.1 dex. 
The values of total \jstar\ calculated for our sample are listed in \autoref{tab:sample} (column 10).

\begin{figure}
	\includegraphics[width=0.49\textwidth]{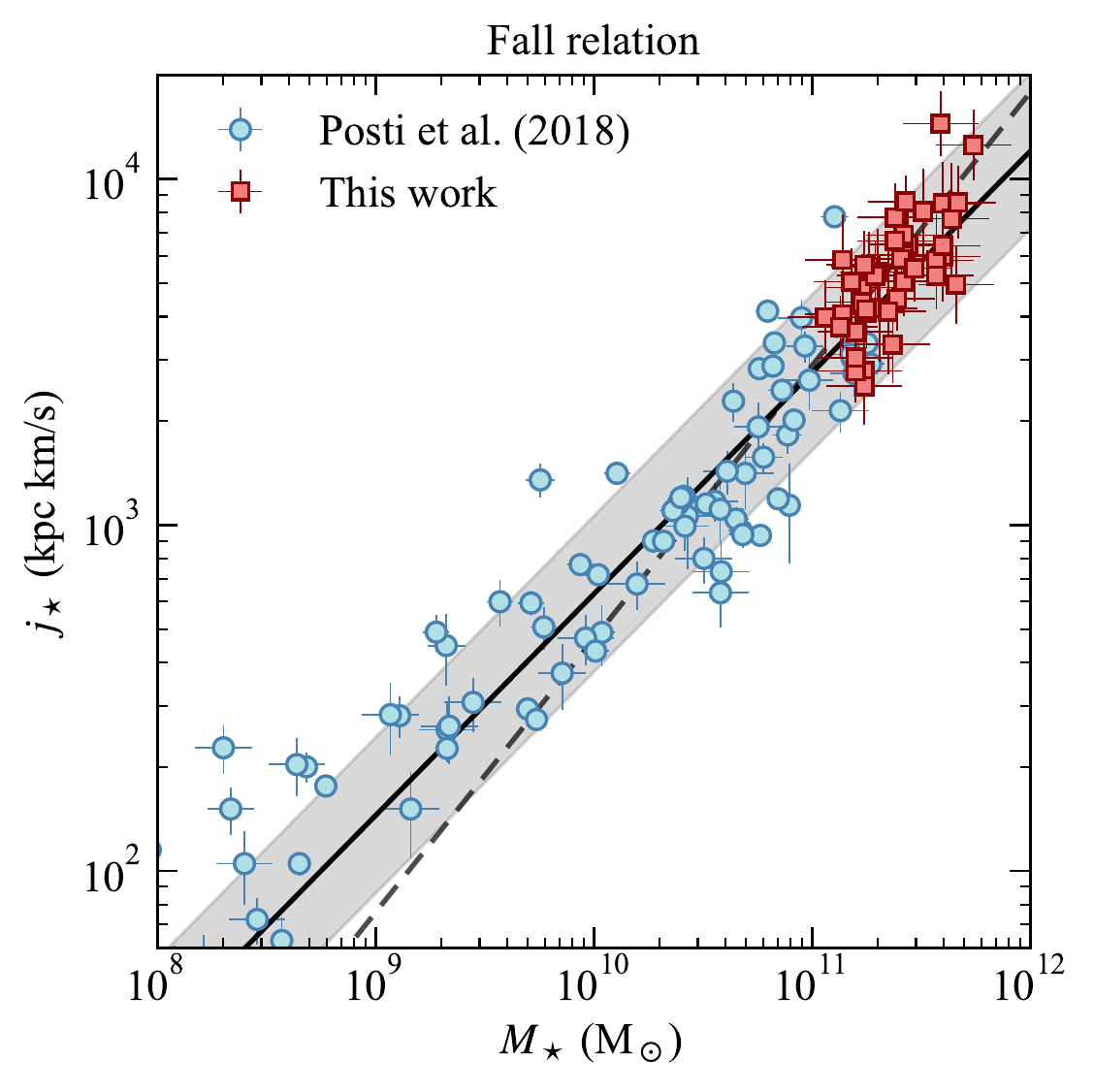}
    \caption{Fall relation ($\jstar-\mstar$) for massive spiral galaxies.
    Our sample is shown with red squares, cyan circles denote galaxies from the SPARC catalog \citep{Posti+2018b}.
    Solid and dashed black lines are linear fits in the form $\log \jstar = \gamma\log \mstar + \delta$ to all galaxies ($\gamma=0.64$, $\delta=-3.60$) and to galaxies with $\log(\mstar/\mo)\gtrsim10.5$ ($\gamma=0.79$, $\delta=-5.23$), respectively. The grey band marks the orthogonal scatter $\sigma_\bot=0.19$ dex.}
    \label{fig:Fall}
\end{figure}

The Fall relation for our massive disc galaxies is presented in \autoref{fig:Fall} (red squares). 
We compare it with the Fall relation at lower stellar masses determined by \citet{Posti+2018b} using galaxies from the SPARC sample (cyan circles). 
The solid black line indicates the best-fit relation $\log\jstar = (0.64\pm0.11)\log\mstar - (3.60\pm0.21) $ to the entire sample, with an orthogonal scatter of 0.19 dex (grey-shaded region), in good agreement with previous determinations \citep{Fall+1983, Fall+2013, Obreschkow+2014,Cortese+2016,Fall+2018, Posti+2018b,Murugeshan+2020, ManceraPina+2021,ManceraPina+2021b}. 
A slope of 0.64 is very close to the value of 2/3 predicted by tidal torque theories, indicating that $f_j$ must be almost precisely a constant.
The dashed line is the best-fit to galaxies with $\log(\mstar/\mo) \geq 10.5$, $\log\jstar = (0.79\pm0.13)\log\mstar - (5.23\pm0.23)$.
As for the Tully-Fisher relations, our massive galaxies are broadly consistent with being on a single, unbroken relation, with the only significant outlier being, again, one of the two fastest rotators (2MFGC08638).
We note that most of our galaxies lie slightly above the linear relation, which may suggest the beginning of a gradual curvature of the relation at the high-mass end. 
However, this slight bend toward larger \jstar\ might be partially due to our secondary sample selection criterion (Section~\ref{sec:data}), which favored galaxies with large disc sizes, hence with larger specific angular momenta ($\jstar \propto R_\mathrm{d}\vflat$).


\section{Discussion}
\label{sec:discussion}

Some previous studies have suggested that there might be a bend in the Tully-Fisher relation at the high-mass end \citep[e.g.,][]{Peletier+1993,Verheijen+2001,DeRijcke+2007,Noordermeer+2007b,Courtois+2015,Ogle+2019}.
Most of these investigations focus on early-type or lenticular galaxies, which seem to lie on a relation that is offset from that of spiral galaxies \citep[e.g.][]{Williams+2010,Cortesi+2013} and/or appear to break at high-mass regime \citep[e.g.,][however, see also \citealt{denHeijer+2015}]{Davis+2016}. 
The Tully-Fisher relation for massive discs is less certain, due to the scarcity of these systems.
\citet{Noordermeer+2007b} used a sample of 48 fairly massive galaxies ($\mstar\lesssim2\times10^{11} \, \mo$) with extended \hi\ rotation curves and found that late-type discs \citep{Spekkens+2006} always lie on the stellar Tully-Fisher relation determined by less massive galaxies, while earlier-type discs \citep{Noordermeer+2007} seem to deviate from it.
However, they also showed that this discrepancy is largely reduced when the asymptotic (flat) velocity is used instead of other velocity proxies, because earlier-type galaxies often have centrally peaked rotation curves that decline at large radii. 
The discrepancy disappears when using baryonic mass instead of stellar mass in the Tully-Fisher relation. 
Our galaxy sample, which extends to higher masses than that of \cite{Noordermeer+2007b} and on average includes later-type discs with fairly flat rotation curves, confirms that the stellar and baryonic Tully-Fisher relations extend up to $\mstar \simeq 6\times 10^{11} \, \mo$ as unbroken power laws. 

We use the analogous $M_{\rm h}-V_{\rm h}$ and $j_{\rm h}-M_{\rm h}$ relations for dark matter halos as a reference for the observed Tully-Fisher and Fall relations, and we introduce the stellar mass fraction $f_M=\mstar/\Mh$, the angular momentum fraction $f_j = \jstar/\jh$ and the ratio $f_V=\vflat/\Vh$. 
With these three fractions, the Tully-Fisher and Fall relations become $\mstar\propto f_M (\vflat/f_V)^3$ and $\jstar\propto f_j (\mstar/f_M)^{2/3}$.
We use the statistical method of \cite{Posti+2019}, on a galaxy sample consisting of their original sample plus our \galnum\ massive spirals, to determine how the fractions $f_M$, $f_j$, and $f_V$ vary with mass and to test for the presence of bends in the observed scaling relations. 
The results of our new analysis are perfectly compatible with those of \cite{Posti+2019}.
In particular, we find that the data statistically prefer a model in which $f_M$ is approximately proportional to $\mstar^{1/4}$, while both the angular momentum fraction $f_j$ and the velocity fraction $f_V$ are roughly constant with mass ($f_j\sim 0.7$, $f_V\sim 1$). 
Thus, on average, galactic discs have almost the same specific angular momentum as their dark matter halos, independent of mass. This result is consistent with early models of disc formation \citep[e.g.][]{Fall+1980,Dalcanton+1997,Mo+98}, although the physical mechanisms responsible for it have not yet been fully elucidated \citep{Romanowsky+2012,DeFelippis+2017,Grand+2019}.
On the other hand, a model with a prior imposed on $f_M$ that follows the standard stellar-to-halo mass relation with turnover \citep[e.g.][]{Wechsler+2018} is highly disfavoured by the observed scaling laws.
Our statistical analysis confirms that $f_M$ increases as a power law of $\mstar$ for spiral galaxies up to the highest masses that we can measure, while, for lenticular and elliptical galaxies, $f_M$ turns over at $\mstar \sim 5\times 10^{10}\mo$ \citep{Posti+2019b,PostiFall21}.

\citet{Ogle+2019}, using a sub-sample of the galaxies analysed in the present work (see \autoref{tab:sample}), reported that the baryonic Tully-Fisher relation and the Fall relation break above a critical stellar mass $\log(\mstar/\mo) \simeq 11.5$, which seems to be in contrast with our findings.
In this paper, we analyse a galaxy sample quite larger than \citeauthor{Ogle+2019}'s (\galnum\ vs 23), and we take advantage of a more sophisticated approach to derive galaxy kinematics. 
While we obtain rotation curves through a full modelling of the long-slit observations and we assume the $\vflat$ as velocity estimator for the Tully-Fisher relations, \citeauthor{Ogle+2019} used the maximum rotation speed $V_\mathrm{max}$ measured directly from the flux-weighted centroid wavelength of the \ha\ emission line.
We stress that, although for some galaxies the $V_\mathrm{max}$ estimated by \citet{Ogle+2019} is slightly larger than our $\vflat$, for most galaxies their velocities are consistent with our within the errors.
For the Fall relation, we calculated the specific angular momenta from the converged profiles $\jstar(R)$, while \citeauthor{Ogle+2019} used the estimator $\jstar=2V_\mathrm{max} R_\mathrm{d}$ for pure exponential discs with scale radius $R_\mathrm{d}$. 
Our $\jstar(R)$ profiles on average converge to the estimator for exponential discs quite well, as shown in \autoref{fig:jprof}. 
However, we found that the scale radii used in \citeauthor{Ogle+2019}, which were taken from automated SDSS bulge-disc decomposition with fixed Sérsic index by \citet{Simard+2011}, are systematically larger than the values estimated here. 
This results in their estimates of $\jstar$ being typically larger than ours.
\citet{Ogle+2019}'s claim of a break in the scaling relations was mostly based on 6 outlier galaxies with very high rotation velocities.
We had to exclude two of these galaxies (OGC1304 and OGC0139) from our sample because a more attentive analysis showed that their current \ha\ data are not good enough to determine an unambiguous rotation velocity, due to either asymmetric (OGC1304) or anomalous (OGC0139) \ha\ emission (see \autoref{fig:discarded}). 
For two other galaxies, OGC0441 and 2MASXJ1123, we derived significantly smaller rotation speeds than \citeauthor{Ogle+2019}'s (384 vs 444 $\kms$ and 386 vs 436 $\kms$, respectively), and our measurements move these galaxies back onto the linear relations. 
In the case of OGC0441, the velocity discrepancy is mainly attributable to the different inclination angle, $44\de$ estimated in this work compared to the $39\de$ assumed by \citet{Ogle+2019}.
The \ha\ emission of 2MASXJ1123 is quite asymmetric, with the approaching side being less extended than the receding side and seemingly reaching higher velocities (see second row of \autoref{fig:maps}); this might have affected the estimate of the systemic and maximum velocity in \citet{Ogle+2019}.
The last two galaxies, 2MFGC08638 and 2MFGC12344, are still outliers in the scaling relations determined in this paper, in agreement with \citeauthor{Ogle+2019}'s findings.
However, we did not find evidence of any further extremely fast-rotating disc ($\vrot\gtrsim400 \, \kms$) amongst the 20 massive spirals that we added to \citet{Ogle+2019}'s galaxy sample.
Our analysis therefore leaves us with only two rapid rotators that appear to lie outside the Tully-Fisher and Fall relations, which is insufficient to test the claim that there is a break at $\vrot>400 \, \kms$.

Massive discs are rare in the local Universe and fast rotators appear to be even rarer.
Our current sample of \galnum\ massive discs revealed only 3 galaxies with $\vflat>400$ $\kms$, not enough to determine the shapes of the scaling relations in this range of extreme rotation velocities with any confidence.
To detect or refute a bend at a statistically significant level, we would likely need $\sim10-20$ galaxies with $\vflat>400$ $\kms$.  
It might seem, at first glance, that this could be achieved with a relatively modest extension of the observing program described in this work. 
Unfortunately, however, it is difficult to know which galaxies are fast rotators until their rotation curves are measured.  
Because the space density of galaxies falls so rapidly with luminosity, size and \hi\ line width, it is necessary to observe many more galaxies that might have $\vflat>400$ $\kms$ for every one that actually does exceed this limit.  
In practice, this means measuring the rotation curves of all the galaxies in a sample with selection criteria similar to those of the present sample, but $\sim5-10$ times larger, amounting to $\sim200-400$ galaxies. While not impossible, this would be a major undertaking.

The two galaxies in our sample with the highest rotational speeds, 2MFGC08638 and 2MFGC12344, are the only significant outliers in the scaling relations.
The extreme rotation velocities ($> 450 \, \kms$) of these two galaxies place them on the high-\vflat\ side of the Tully-Fisher relations (\autoref{fig:TF}) and the high-\jstar\ side of the Fall relation (\autoref{fig:Fall}).
They have also exceptionally large stellar discs that extend up to 200 kpc in diameter in $z$ band.
These two galaxies are hard to reconcile with the relations determined at lower masses: either a significantly larger stellar mass or a significantly smaller asymptotic velocity would be required.
Both these galaxies are highly inclined systems and we may expect some unaccounted dust absorption even at 3.4 $\upmu$m, which may imply a $\mstar/L_\mathrm{W1}$ larger than the assumed value of 0.6.
A mass-to-light ratio of $\mstar/L_\mathrm{W1}\simeq1.5$ would take these two galaxies back on the Tully-Fisher and Fall relations. 
Although such a high $\mstar/L_\mathrm{W1}$ can not be completely ruled out on the basis of stellar population synthesis models \citep[][]{Into+2013,Fall+2013}, it would imply stellar masses larger than $10^{12} \, \mo$, which would suspiciously make these two galaxies the most massive discs ever known. 
Another possibility is that these galaxies have rotation curves that sharply decline beyond the optical radius: a 20-30\% decrease in the asymptotic velocity would make them compatible with the Tully-Fisher relation. 
However, this kind of extreme decline has been rarely seen even in massive early-type discs \citep[e.g.,][]{Noordermeer+2007}, and it seems unlikely to happen in late-type discs like 2MFGC08638 and 2MFGC12344.
We also note that we ruled out the possibility of a scenario where the high-velocity \ha\ emission is due to a peripheral ring of gas rather than to a filled galactic disc (see Section~\ref{sec:rotcurs}).
\hi\ data of the extended gas disc would be helpful to constrain better the asymptotic velocity of these two massive spiral galaxies. 

Our findings suggest that most extremely-massive spiral galaxies are following an evolutionary path similar to that of less massive spirals. 
These galaxies must have been continuing to assemble their stellar mass through secular evolution and minor mergers. 
Their current SFRs are of the order of $10-20 \, \moyr$, which implies that these systems may have easily built stellar masses larger than $10^{11} \, \mo$ throughout cosmic time, if we assume that, similarly to less massive spirals, they have been forming stars at a nearly-constant or declining rate \citep{Panter+2007,Madau+2014}.
These galaxies must have experienced a relatively quiet evolution, without undergoing any major disruptive event, such as major mergers or strong AGN feedback, capable of transforming their disc-like morphology and of moving them off the scaling relations. 
Finally, our analysis indicates that star-forming galaxies have stellar/baryonic masses that scale with the dark matter halos in a self-similar way up to the very high-mass end.


\section{Conclusions}
\label{sec:conc}
In this work, we used new \ha-\nii\ long-slit observations to study the kinematics of \galnum\ extremely massive spiral galaxies with $\log (\mstar / \mo) > 11$. 
We constructed 3D kinematic models of rotating discs, we extracted a long-slit mock observation from them and fitted it to our emission-line observations.
This method allows us to deal with well-known biases due to instrumental response (PSF and spectrograph resolution) and to the high inclination angles of several galaxies in our sample, returning accurate rotation velocities and gas velocity dispersions.
All galaxies analysed show rotation curves that flatten within the optical radius, velocity dispersions of $25-30\, \kms$, typical of the ionized gas component, and $\vrot/\vdisp\gg 10$.
Two galaxies in our sample (2MFGC08638 and 2MFG12344) are fast rotators ($\vflat > 450 \, \kms$), with the most extreme galaxy, 2MFG12344, reaching an impressive rotation velocity of $\simeq 530 \, \kms$ in the outer regions of the optical disc.

From the derived kinematics, we built the stellar and baryonic Tully-Fisher relations and the Fall relation. 
To be consistent with the latest determinations of these scaling laws from \hi\ observations, we used the flat part of the rotation curve as velocity indicator for the Tully-Fisher relation and we estimated the total stellar specific angular momentum from the converged $\jstar(R)$ profiles. 
In contrast with some previous results, we found that massive discs lie on both the Tully-Fisher and the Fall relations determined for less massive spiral galaxies, without any strong evidence for a bend or break at the high-mass end.
The only marginal outliers are the two fastest rotators, which might be difficult to reconcile with the known scaling relations.
While a larger population of such outliers could exist at $\log (\mstar / \mo) > 11.5$, with the data currently available we ruled out the presence of a break at the high-mass end of the Tully-Fisher and Fall relations.
The kinematics and scaling relations derived in this paper indicate that most of these giant discs are scaled up versions of regular spiral galaxies.

\section*{Acknowledgements}

E.M.D.T. was supported by the US National Science Foundation under grant 1616177.
L.P. acknowledges support from the Centre National d'\'{E}tudes Spatiales (CNES) and from the European Research Council (ERC) under the European Unions Horizon 2020 research and innovation program (grant agreement No. 834148).
This paper utilizes data from the Southern African Large Telescope (SALT), the Hale Telescope at Palomar Observatory, the Wide-field Infrared Survey Explorer (WISE), the Sloan Digital Sky Survey (SDSS), and the Legacy Surveys. 
WISE is a joint project of the University of California, Los Angeles, and the Jet Propulsion Laboratory/California Institute of Technology, funded by NASA. 
Funding for the Sloan Digital Sky Survey IV has been provided by the Alfred P.\ Sloan Foundation, the U.\S.\ Department of Energy Office of Science, and the Participating Institutions. The Legacy Surveys consist of three individual and complementary projects: the Dark Energy Camera Legacy Survey (DECaLS), the Beijing-Arizona Sky Survey (BASS), and the Mayall z-band Legacy Survey (MzLS), conducted using facilities supported by NSF's OIR Lab, which is operated by the Association of Universities for Research in Astronomy (AURA) under a cooperative agreement with the National Science Foundation.
This research made use of Photutils, an Astropy package for detection and photometry of astronomical sources.
This work relied on the NASA/IPAC Extragalactic Database, operated by the Jet Propulsion Laboratory, California Institute of Technology, under contract with NASA.

\section*{Data availability}
The data underlying this article will be shared on reasonable request to the corresponding author.


\bibliographystyle{mnras}
\bibliography{biblio_ss}

\begin{thebibliography}{}
\makeatletter
\relax
\def\mn@urlcharsother{\let\do\@makeother \do\$\do\&\do\#\do\^\do\_\do\%\do\~}
\def\mn@doi{\begingroup\mn@urlcharsother \@ifnextchar [ {\mn@doi@}
  {\mn@doi@[]}}
\def\mn@doi@[#1]#2{\def\@tempa{#1}\ifx\@tempa\@empty \href
  {http://dx.doi.org/#2} {doi:#2}\else \href {http://dx.doi.org/#2} {#1}\fi
  \endgroup}
\def\mn@eprint#1#2{\mn@eprint@#1:#2::\@nil}
\def\mn@eprint@arXiv#1{\href {http://arxiv.org/abs/#1} {{\tt arXiv:#1}}}
\def\mn@eprint@dblp#1{\href {http://dblp.uni-trier.de/rec/bibtex/#1.xml}
  {dblp:#1}}
\def\mn@eprint@#1:#2:#3:#4\@nil{\def\@tempa {#1}\def\@tempb {#2}\def\@tempc
  {#3}\ifx \@tempc \@empty \let \@tempc \@tempb \let \@tempb \@tempa \fi \ifx
  \@tempb \@empty \def\@tempb {arXiv}\fi \@ifundefined
  {mn@eprint@\@tempb}{\@tempb:\@tempc}{\expandafter \expandafter \csname
  mn@eprint@\@tempb\endcsname \expandafter{\@tempc}}}

\bibitem[\protect\citeauthoryear{{Andersen}, {Bershady}, {Sparke}, {Gallagher},
  {Wilcots}, {van Driel}  \& {Monnier-Ragaigne}}{{Andersen}
  et~al.}{2006}]{Andersen+2006}
{Andersen} D.~R.,  {Bershady} M.~A.,  {Sparke} L.~S.,  {Gallagher} John~S. I.,
  {Wilcots} E.~M.,  {van Driel} W.,   {Monnier-Ragaigne} D.,  2006, \mn@doi
  [\apjs] {10.1086/506609}, \href
  {https://ui.adsabs.harvard.edu/abs/2006ApJS..166..505A} {166, 505}

\bibitem[\protect\citeauthoryear{{Astropy Collaboration} et~al.,}{{Astropy
  Collaboration} et~al.}{2013}]{astropy:2013}
{Astropy Collaboration} et~al., 2013, \mn@doi [\aap]
  {10.1051/0004-6361/201322068}, \href
  {http://adsabs.harvard.edu/abs/2013A%26A...558A..33A} {558, A33}

\bibitem[\protect\citeauthoryear{{Astropy Collaboration} et~al.,}{{Astropy
  Collaboration} et~al.}{2018}]{astropy:2018}
{Astropy Collaboration} et~al., 2018, \mn@doi [aj] {10.3847/1538-3881/aabc4f},
  \href {https://ui.adsabs.harvard.edu/abs/2018AJ....156..123A} {156, 123}

\bibitem[\protect\citeauthoryear{{Begeman}}{{Begeman}}{1987}]{Begeman+1987}
{Begeman} K.~G.,  1987, PhD thesis, -

\bibitem[\protect\citeauthoryear{{Bekerait{\.{e}}} et~al.,}{{Bekerait{\.{e}}}
  et~al.}{2016}]{Bekeraite+2016}
{Bekerait{\.{e}}} S.,  et~al., 2016, \mn@doi [\apjl]
  {10.3847/2041-8205/827/2/L36}, \href
  {https://ui.adsabs.harvard.edu/abs/2016ApJ...827L..36B} {827, L36}

\bibitem[\protect\citeauthoryear{{Bloom} et~al.,}{{Bloom}
  et~al.}{2017}]{Bloom+2017}
{Bloom} J.~V.,  et~al., 2017, \mn@doi [\mnras] {10.1093/mnras/stx1701}, \href
  {https://ui.adsabs.harvard.edu/abs/2017MNRAS.472.1809B} {472, 1809}

\bibitem[\protect\citeauthoryear{{Bradley}, {Sipocz}, {Robitaille}, Erik  \& et
  al.}{{Bradley} et~al.}{2020}]{larry_bradley_2020_4044744}
{Bradley} L.,  {Sipocz} B.,  {Robitaille} T.,  Erik T.,   et al. 2020,
  astropy/photutils: 1.0.0, \mn@doi{10.5281/zenodo.4044744}, \url
  {https://doi.org/10.5281/zenodo.4044744}

\bibitem[\protect\citeauthoryear{{Buckley}, {Swart}  \& {Meiring}}{{Buckley}
  et~al.}{2006}]{Buckley+2006}
{Buckley} D. A.~H.,  {Swart} G.~P.,   {Meiring} J.~G.,  2006, {Completion and
  commissioning of the Southern African Large Telescope}.
p. 62670Z, \mn@doi{10.1117/12.673750}

\bibitem[\protect\citeauthoryear{{Bullock}, {Dekel}, {Kolatt}, {Kravtsov},
  {Klypin}, {Porciani}  \& {Primack}}{{Bullock} et~al.}{2001}]{Bullock+2001}
{Bullock} J.~S.,  {Dekel} A.,  {Kolatt} T.~S.,  {Kravtsov} A.~V.,  {Klypin}
  A.~A.,  {Porciani} C.,   {Primack} J.~R.,  2001, \mn@doi [\apj]
  {10.1086/321477}, \href
  {https://ui.adsabs.harvard.edu/abs/2001ApJ...555..240B} {555, 240}

\bibitem[\protect\citeauthoryear{{Burgh}, {Nordsieck}, {Kobulnicky},
  {Williams}, {O'Donoghue}, {Smith}  \& {Percival}}{{Burgh}
  et~al.}{2003}]{Burgh+2003}
{Burgh} E.~B.,  {Nordsieck} K.~H.,  {Kobulnicky} H.~A.,  {Williams} T.~B.,
  {O'Donoghue} D.,  {Smith} M.~P.,   {Percival} J.~W.,  2003, {Prime Focus
  Imaging Spectrograph for the Southern African Large Telescope: optical
  design}.
pp 1463--1471, \mn@doi{10.1117/12.460312}

\bibitem[\protect\citeauthoryear{{Cattaneo}, {Salucci}  \&
  {Papastergis}}{{Cattaneo} et~al.}{2014}]{Cattaneo+14}
{Cattaneo} A.,  {Salucci} P.,   {Papastergis} E.,  2014, \mn@doi [\apj]
  {10.1088/0004-637X/783/2/66}, \href
  {https://ui.adsabs.harvard.edu/abs/2014ApJ...783...66C} {783, 66}

\bibitem[\protect\citeauthoryear{{Cluver} et~al.,}{{Cluver}
  et~al.}{2014}]{Cluver+2014}
{Cluver} M.~E.,  et~al., 2014, \mn@doi [\apj] {10.1088/0004-637X/782/2/90},
  \href {https://ui.adsabs.harvard.edu/abs/2014ApJ...782...90C} {782, 90}

\bibitem[\protect\citeauthoryear{{Cluver}, {Jarrett}, {Dale}, {Smith}, {August}
   \& {Brown}}{{Cluver} et~al.}{2017}]{Cluver+2017}
{Cluver} M.~E.,  {Jarrett} T.~H.,  {Dale} D.~A.,  {Smith} J. D.~T.,  {August}
  T.,   {Brown} M.~J.~I.,  2017, \mn@doi [\apj] {10.3847/1538-4357/aa92c7},
  \href {https://ui.adsabs.harvard.edu/abs/2017ApJ...850...68C} {850, 68}

\bibitem[\protect\citeauthoryear{{Cortese} et~al.,}{{Cortese}
  et~al.}{2016}]{Cortese+2016}
{Cortese} L.,  et~al., 2016, \mn@doi [MNRAS] {10.1093/mnras/stw1891}, \href
  {https://ui.adsabs.harvard.edu/abs/2016MNRAS.463..170C} {463, 170}

\bibitem[\protect\citeauthoryear{{Cortesi} et~al.,}{{Cortesi}
  et~al.}{2013}]{Cortesi+2013}
{Cortesi} A.,  et~al., 2013, \mn@doi [\mnras] {10.1093/mnras/stt529}, \href
  {https://ui.adsabs.harvard.edu/abs/2013MNRAS.432.1010C} {432, 1010}

\bibitem[\protect\citeauthoryear{{Courtois}, {Zaritsky}, {Sorce}  \&
  {Pomar{\`e}de}}{{Courtois} et~al.}{2015}]{Courtois+2015}
{Courtois} H.~M.,  {Zaritsky} D.,  {Sorce} J.~G.,   {Pomar{\`e}de} D.,  2015,
  \mn@doi [\mnras] {10.1093/mnras/stv071}, \href
  {https://ui.adsabs.harvard.edu/abs/2015MNRAS.448.1767C} {448, 1767}

\bibitem[\protect\citeauthoryear{{Dalcanton}, {Spergel}  \&
  {Summers}}{{Dalcanton} et~al.}{1997}]{Dalcanton+1997}
{Dalcanton} J.~J.,  {Spergel} D.~N.,   {Summers} F.~J.,  1997, \mn@doi [ApJ]
  {10.1086/304182}, \href
  {https://ui.adsabs.harvard.edu/abs/1997ApJ...482..659D} {482, 659}

\bibitem[\protect\citeauthoryear{{Davis}, {Greene}, {Ma}, {Pand ya},
  {Blakeslee}, {McConnell}  \& {Thomas}}{{Davis} et~al.}{2016}]{Davis+2016}
{Davis} T.~A.,  {Greene} J.,  {Ma} C.-P.,  {Pand ya} V.,  {Blakeslee} J.~P.,
  {McConnell} N.,   {Thomas} J.,  2016, \mn@doi [\mnras]
  {10.1093/mnras/stv2313}, \href
  {https://ui.adsabs.harvard.edu/abs/2016MNRAS.455..214D} {455, 214}

\bibitem[\protect\citeauthoryear{{De Rijcke}, {Zeilinger}, {Hau}, {Prugniel}
  \& {Dejonghe}}{{De Rijcke} et~al.}{2007}]{DeRijcke+2007}
{De Rijcke} S.,  {Zeilinger} W.~W.,  {Hau} G. K.~T.,  {Prugniel} P.,
  {Dejonghe} H.,  2007, \mn@doi [\apj] {10.1086/512717}, \href
  {https://ui.adsabs.harvard.edu/abs/2007ApJ...659.1172D} {659, 1172}

\bibitem[\protect\citeauthoryear{{DeFelippis}, {Genel}, {Bryan}  \&
  {Fall}}{{DeFelippis} et~al.}{2017}]{DeFelippis+2017}
{DeFelippis} D.,  {Genel} S.,  {Bryan} G.~L.,   {Fall} S.~M.,  2017, \mn@doi
  [ApJ] {10.3847/1538-4357/aa6dfc}, \href
  {https://ui.adsabs.harvard.edu/abs/2017ApJ...841...16D} {841, 16}

\bibitem[\protect\citeauthoryear{{Dekel} \& {Silk}}{{Dekel} \&
  {Silk}}{1986}]{Dekel+1986}
{Dekel} A.,  {Silk} J.,  1986, \mn@doi [\apj] {10.1086/164050}, \href
  {https://ui.adsabs.harvard.edu/abs/1986ApJ...303...39D} {303, 39}

\bibitem[\protect\citeauthoryear{{Dey} et~al.,}{{Dey} et~al.}{2019}]{Dey+2019}
{Dey} A.,  et~al., 2019, \mn@doi [\aj] {10.3847/1538-3881/ab089d}, \href
  {https://ui.adsabs.harvard.edu/abs/2019AJ....157..168D} {157, 168}

\bibitem[\protect\citeauthoryear{{Di Teodoro} \& {Fraternali}}{{Di Teodoro} \&
  {Fraternali}}{2015}]{DiTeodoro&Fraternali15}
{Di Teodoro} E.~M.,  {Fraternali} F.,  2015, \mn@doi [\mnras]
  {10.1093/mnras/stv1213}, \href
  {https://ui.adsabs.harvard.edu/abs/2015MNRAS.451.3021D} {451, 3021}

\bibitem[\protect\citeauthoryear{{Di Teodoro} et~al.,}{{Di Teodoro}
  et~al.}{2018}]{DiTeodoro+2018}
{Di Teodoro} E.~M.,  et~al., 2018, \mn@doi [\mnras] {10.1093/mnras/sty175},
  \href {https://ui.adsabs.harvard.edu/abs/2018MNRAS.476..804D} {476, 804}

\bibitem[\protect\citeauthoryear{{Dutton} \& {van den Bosch}}{{Dutton} \& {van
  den Bosch}}{2012}]{DuttonvdBosch12}
{Dutton} A.~A.,  {van den Bosch} F.~C.,  2012, \mn@doi [\mnras]
  {10.1111/j.1365-2966.2011.20339.x}, \href
  {https://ui.adsabs.harvard.edu/abs/2012MNRAS.421..608D} {421, 608}

\bibitem[\protect\citeauthoryear{{Eisenstein} et~al.,}{{Eisenstein}
  et~al.}{2011}]{Eisenstein+2011}
{Eisenstein} D.~J.,  et~al., 2011, \mn@doi [\aj] {10.1088/0004-6256/142/3/72},
  \href {https://ui.adsabs.harvard.edu/abs/2011AJ....142...72E} {142, 72}

\bibitem[\protect\citeauthoryear{{Fall}}{{Fall}}{1983}]{Fall+1983}
{Fall} S.~M.,  1983, in {Athanassoula} E.,  ed., ~ Vol. 100, Internal
  Kinematics and Dynamics of Galaxies. pp 391--398

\bibitem[\protect\citeauthoryear{{Fall} \& {Efstathiou}}{{Fall} \&
  {Efstathiou}}{1980}]{Fall+1980}
{Fall} S.~M.,  {Efstathiou} G.,  1980, \mn@doi [\mnras]
  {10.1093/mnras/193.2.189}, \href
  {https://ui.adsabs.harvard.edu/abs/1980MNRAS.193..189F} {193, 189}

\bibitem[\protect\citeauthoryear{{Fall} \& {Romanowsky}}{{Fall} \&
  {Romanowsky}}{2013}]{Fall+2013}
{Fall} S.~M.,  {Romanowsky} A.~J.,  2013, \mn@doi [\apjl]
  {10.1088/2041-8205/769/2/L26}, \href
  {https://ui.adsabs.harvard.edu/abs/2013ApJ...769L..26F} {769, L26}

\bibitem[\protect\citeauthoryear{{Fall} \& {Romanowsky}}{{Fall} \&
  {Romanowsky}}{2018}]{Fall+2018}
{Fall} S.~M.,  {Romanowsky} A.~J.,  2018, \mn@doi [\apj]
  {10.3847/1538-4357/aaeb27}, \href
  {https://ui.adsabs.harvard.edu/abs/2018ApJ...868..133F} {868, 133}

\bibitem[\protect\citeauthoryear{{Ferrero} et~al.,}{{Ferrero}
  et~al.}{2017}]{Ferrero+2017}
{Ferrero} I.,  et~al., 2017, \mn@doi [\mnras] {10.1093/mnras/stw2691}, \href
  {https://ui.adsabs.harvard.edu/abs/2017MNRAS.464.4736F} {464, 4736}

\bibitem[\protect\citeauthoryear{{Fouque}, {Bottinelli}, {Gouguenheim}  \&
  {Paturel}}{{Fouque} et~al.}{1990}]{Fouque+1990}
{Fouque} P.,  {Bottinelli} L.,  {Gouguenheim} L.,   {Paturel} G.,  1990,
  \mn@doi [\apj] {10.1086/168288}, \href
  {https://ui.adsabs.harvard.edu/abs/1990ApJ...349....1F} {349, 1}

\bibitem[\protect\citeauthoryear{{Fraternali}, {Sancisi}  \&
  {Kamphuis}}{{Fraternali} et~al.}{2011}]{Fraternali+11}
{Fraternali} F.,  {Sancisi} R.,   {Kamphuis} P.,  2011, \mn@doi [\aap]
  {10.1051/0004-6361/201116634}, \href
  {https://ui.adsabs.harvard.edu/abs/2011A&A...531A..64F} {531, A64}

\bibitem[\protect\citeauthoryear{{Giovanelli}, {Haynes}, {Rubin}  \&
  {Ford}}{{Giovanelli} et~al.}{1986}]{Giovanelli+1986}
{Giovanelli} R.,  {Haynes} M.~P.,  {Rubin} V.~C.,   {Ford} W.~K. J.,  1986,
  \mn@doi [\apjl] {10.1086/184613}, \href
  {https://ui.adsabs.harvard.edu/abs/1986ApJ...301L...7G} {301, L7}

\bibitem[\protect\citeauthoryear{{Grand} et~al.,}{{Grand}
  et~al.}{2019}]{Grand+2019}
{Grand} R. J.~J.,  et~al., 2019, \mn@doi [MNRAS] {10.1093/mnras/stz2928}, \href
  {https://ui.adsabs.harvard.edu/abs/2019MNRAS.490.4786G} {490, 4786}

\bibitem[\protect\citeauthoryear{Huyer \& Neumaier}{Huyer \&
  Neumaier}{1998}]{Huyer&Neumaier98}
Huyer W.,  Neumaier A.,  1998, \mn@doi [Journal of Global Optimization]
  {10.1023/A:1008382309369}, 14

\bibitem[\protect\citeauthoryear{{Into} \& {Portinari}}{{Into} \&
  {Portinari}}{2013}]{Into+2013}
{Into} T.,  {Portinari} L.,  2013, \mn@doi [\mnras] {10.1093/mnras/stt071},
  \href {https://ui.adsabs.harvard.edu/abs/2013MNRAS.430.2715I} {430, 2715}

\bibitem[\protect\citeauthoryear{{Jarrett}, {Chester}, {Cutri}, {Schneider},
  {Skrutskie}  \& {Huchra}}{{Jarrett} et~al.}{2000}]{Jarrett+2000}
{Jarrett} T.~H.,  {Chester} T.,  {Cutri} R.,  {Schneider} S.,  {Skrutskie} M.,
   {Huchra} J.~P.,  2000, \mn@doi [\aj] {10.1086/301330}, \href
  {https://ui.adsabs.harvard.edu/abs/2000AJ....119.2498J} {119, 2498}

\bibitem[\protect\citeauthoryear{{Jarrett}, {Cluver}, {Brown}, {Dale}, {Tsai}
  \& {Masci}}{{Jarrett} et~al.}{2019}]{Jarrett+2019}
{Jarrett} T.~H.,  {Cluver} M.~E.,  {Brown} M.~J.~I.,  {Dale} D.~A.,  {Tsai}
  C.~W.,   {Masci} F.,  2019, \mn@doi [\apjs] {10.3847/1538-4365/ab521a}, \href
  {https://ui.adsabs.harvard.edu/abs/2019ApJS..245...25J} {245, 25}

\bibitem[\protect\citeauthoryear{{Jedrzejewski}}{{Jedrzejewski}}{1987}]{Jedrzejewski+1987}
{Jedrzejewski} R.~I.,  1987, \mn@doi [\mnras] {10.1093/mnras/226.4.747}, \href
  {https://ui.adsabs.harvard.edu/abs/1987MNRAS.226..747J} {226, 747}

\bibitem[\protect\citeauthoryear{{Karachentsev}, {Mitronova}, {Karachentseva},
  {Kudrya}  \& {Jarrett}}{{Karachentsev} et~al.}{2002}]{Karachentsev+2002}
{Karachentsev} I.~D.,  {Mitronova} S.~N.,  {Karachentseva} V.~E.,  {Kudrya}
  Y.~N.,   {Jarrett} T.~H.,  2002, \mn@doi [\aap] {10.1051/0004-6361:20021451},
  \href {https://ui.adsabs.harvard.edu/abs/2002A&A...396..431K} {396, 431}

\bibitem[\protect\citeauthoryear{{Kennicutt}}{{Kennicutt}}{1998}]{Kennicutt+1998}
{Kennicutt} Robert~C. J.,  1998, \mn@doi [\apj] {10.1086/305588}, \href
  {https://ui.adsabs.harvard.edu/abs/1998ApJ...498..541K} {498, 541}

\bibitem[\protect\citeauthoryear{{Kent}}{{Kent}}{1987}]{Kent+1987}
{Kent} S.~M.,  1987, \mn@doi [\aj] {10.1086/114366}, \href
  {https://ui.adsabs.harvard.edu/abs/1987AJ.....93..816K} {93, 816}

\bibitem[\protect\citeauthoryear{{Kent}}{{Kent}}{1988}]{Kent+1988}
{Kent} S.~M.,  1988, \mn@doi [\aj] {10.1086/114829}, \href
  {https://ui.adsabs.harvard.edu/abs/1988AJ.....96..514K} {96, 514}

\bibitem[\protect\citeauthoryear{{Lapi}, {Salucci}  \& {Danese}}{{Lapi}
  et~al.}{2018}]{Lapi+2018}
{Lapi} A.,  {Salucci} P.,   {Danese} L.,  2018, \mn@doi [\apj]
  {10.3847/1538-4357/aabf35}, \href
  {https://ui.adsabs.harvard.edu/abs/2018ApJ...859....2L} {859, 2}

\bibitem[\protect\citeauthoryear{{Lelli}, {McGaugh}  \& {Schombert}}{{Lelli}
  et~al.}{2016a}]{Lelli+2016b}
{Lelli} F.,  {McGaugh} S.~S.,   {Schombert} J.~M.,  2016a, \mn@doi [\aj]
  {10.3847/0004-6256/152/6/157}, \href
  {https://ui.adsabs.harvard.edu/abs/2016AJ....152..157L} {152, 157}

\bibitem[\protect\citeauthoryear{{Lelli}, {McGaugh}  \& {Schombert}}{{Lelli}
  et~al.}{2016b}]{Lelli+2016}
{Lelli} F.,  {McGaugh} S.~S.,   {Schombert} J.~M.,  2016b, \mn@doi [\apjl]
  {10.3847/2041-8205/816/1/L14}, \href
  {https://ui.adsabs.harvard.edu/abs/2016ApJ...816L..14L} {816, L14}

\bibitem[\protect\citeauthoryear{{Lelli}, {McGaugh}, {Schombert}, {Desmond}  \&
  {Katz}}{{Lelli} et~al.}{2019}]{Lelli+2019}
{Lelli} F.,  {McGaugh} S.~S.,  {Schombert} J.~M.,  {Desmond} H.,   {Katz} H.,
  2019, \mn@doi [\mnras] {10.1093/mnras/stz205}, \href
  {https://ui.adsabs.harvard.edu/abs/2019MNRAS.484.3267L} {484, 3267}

\bibitem[\protect\citeauthoryear{{Madau} \& {Dickinson}}{{Madau} \&
  {Dickinson}}{2014}]{Madau+2014}
{Madau} P.,  {Dickinson} M.,  2014, \mn@doi [\araa]
  {10.1146/annurev-astro-081811-125615}, \href
  {https://ui.adsabs.harvard.edu/abs/2014ARA&A..52..415M} {52, 415}

\bibitem[\protect\citeauthoryear{{Mancera Pi{\~n}a}, {Posti}, {Fraternali},
  {Adams}  \& {Oosterloo}}{{Mancera Pi{\~n}a} et~al.}{2021a}]{ManceraPina+2021}
{Mancera Pi{\~n}a} P.~E.,  {Posti} L.,  {Fraternali} F.,  {Adams} E. A.~K.,
  {Oosterloo} T.,  2021a, \mn@doi [\aap] {10.1051/0004-6361/202039340}, \href
  {https://ui.adsabs.harvard.edu/abs/2021A&A...647A..76M} {647, A76}

\bibitem[\protect\citeauthoryear{{Mancera Pi{\~n}a}, {Posti}, {Pezzulli},
  {Fraternali}, {Fall}, {Oosterloo}  \& {Adams}}{{Mancera Pi{\~n}a}
  et~al.}{2021b}]{ManceraPina+2021b}
{Mancera Pi{\~n}a} P.~E.,  {Posti} L.,  {Pezzulli} G.,  {Fraternali} F.,
  {Fall} S.~M.,  {Oosterloo} T.,   {Adams} E. A.~K.,  2021b, \mn@doi [A\&A]
  {10.1051/0004-6361/202141574}, \href
  {https://ui.adsabs.harvard.edu/abs/2021A&A...651L..15M} {651, L15}

\bibitem[\protect\citeauthoryear{{Martinsson}, {Verheijen}, {Westfall},
  {Bershady}, {Schechtman-Rook}, {Andersen}  \& {Swaters}}{{Martinsson}
  et~al.}{2013}]{Martinsson+2013}
{Martinsson} T. P.~K.,  {Verheijen} M. A.~W.,  {Westfall} K.~B.,  {Bershady}
  M.~A.,  {Schechtman-Rook} A.,  {Andersen} D.~R.,   {Swaters} R.~A.,  2013,
  \mn@doi [\aap] {10.1051/0004-6361/201220515}, \href
  {https://ui.adsabs.harvard.edu/abs/2013A&A...557A.130M} {557, A130}

\bibitem[\protect\citeauthoryear{{McGaugh}}{{McGaugh}}{2005}]{McGaugh2005}
{McGaugh} S.~S.,  2005, \mn@doi [\apj] {10.1086/432968}, \href
  {https://ui.adsabs.harvard.edu/abs/2005ApJ...632..859M} {632, 859}

\bibitem[\protect\citeauthoryear{{McGaugh}}{{McGaugh}}{2012}]{McGaugh12}
{McGaugh} S.~S.,  2012, \mn@doi [\aj] {10.1088/0004-6256/143/2/40}, \href
  {https://ui.adsabs.harvard.edu/abs/2012AJ....143...40M} {143, 40}

\bibitem[\protect\citeauthoryear{{McGaugh} \& {Schombert}}{{McGaugh} \&
  {Schombert}}{2015}]{McGaugh+2015}
{McGaugh} S.~S.,  {Schombert} J.~M.,  2015, \mn@doi [\apj]
  {10.1088/0004-637X/802/1/18}, \href
  {https://ui.adsabs.harvard.edu/abs/2015ApJ...802...18M} {802, 18}

\bibitem[\protect\citeauthoryear{{McGaugh}, {Schombert}, {Bothun}  \& {de
  Blok}}{{McGaugh} et~al.}{2000}]{McGaugh+2000}
{McGaugh} S.~S.,  {Schombert} J.~M.,  {Bothun} G.~D.,   {de Blok} W.~J.~G.,
  2000, \mn@doi [\apjl] {10.1086/312628}, \href
  {https://ui.adsabs.harvard.edu/abs/2000ApJ...533L..99M} {533, L99}

\bibitem[\protect\citeauthoryear{{Meidt} et~al.,}{{Meidt}
  et~al.}{2014}]{Meidt+2014}
{Meidt} S.~E.,  et~al., 2014, \mn@doi [\apj] {10.1088/0004-637X/788/2/144},
  \href {https://ui.adsabs.harvard.edu/abs/2014ApJ...788..144M} {788, 144}

\bibitem[\protect\citeauthoryear{{Mo}, {Mao}  \& {White}}{{Mo}
  et~al.}{1998}]{Mo+98}
{Mo} H.~J.,  {Mao} S.,   {White} S. D.~M.,  1998, \mn@doi [\mnras]
  {10.1046/j.1365-8711.1998.01227.x}, \href
  {https://ui.adsabs.harvard.edu/abs/1998MNRAS.295..319M} {295, 319}

\bibitem[\protect\citeauthoryear{{Mo}, {van den Bosch}  \& {White}}{{Mo}
  et~al.}{2010}]{MovdBW10}
{Mo} H.,  {van den Bosch} F.~C.,   {White} S.,  2010, {Galaxy Formation and
  Evolution}

\bibitem[\protect\citeauthoryear{{Murugeshan}, {Kilborn}, {Jarrett}, {Wong},
  {Obreschkow}, {Glazebrook}, {Cluver}  \& {Fluke}}{{Murugeshan}
  et~al.}{2020}]{Murugeshan+2020}
{Murugeshan} C.,  {Kilborn} V.,  {Jarrett} T.,  {Wong} O.~I.,  {Obreschkow} D.,
   {Glazebrook} K.,  {Cluver} M.~E.,   {Fluke} C.~J.,  2020, \mn@doi [MNRAS]
  {10.1093/mnras/staa1731}, \href
  {https://ui.adsabs.harvard.edu/abs/2020MNRAS.496.2516M} {496, 2516}

\bibitem[\protect\citeauthoryear{{Naluminsa}, {Elson}  \&
  {Jarrett}}{{Naluminsa} et~al.}{2021}]{Naluminsa+2021}
{Naluminsa} E.,  {Elson} E.~C.,   {Jarrett} T.~H.,  2021, \mn@doi [\mnras]
  {10.1093/mnras/stab067}, \href
  {https://ui.adsabs.harvard.edu/abs/2021MNRAS.tmp..118N} {}

\bibitem[\protect\citeauthoryear{{Navarro} \& {Steinmetz}}{{Navarro} \&
  {Steinmetz}}{2000}]{Navarro+00}
{Navarro} J.~F.,  {Steinmetz} M.,  2000, \mn@doi [\apj] {10.1086/309175}, \href
  {https://ui.adsabs.harvard.edu/abs/2000ApJ...538..477N} {538, 477}

\bibitem[\protect\citeauthoryear{{Noordermeer} \& {Verheijen}}{{Noordermeer} \&
  {Verheijen}}{2007}]{Noordermeer+2007b}
{Noordermeer} E.,  {Verheijen} M.~A.~W.,  2007, \mn@doi [\mnras]
  {10.1111/j.1365-2966.2007.12369.x}, \href
  {https://ui.adsabs.harvard.edu/abs/2007MNRAS.381.1463N} {381, 1463}

\bibitem[\protect\citeauthoryear{{Noordermeer}, {van der Hulst}, {Sancisi},
  {Swaters}  \& {van Albada}}{{Noordermeer} et~al.}{2007}]{Noordermeer+2007}
{Noordermeer} E.,  {van der Hulst} J.~M.,  {Sancisi} R.,  {Swaters} R.~S.,
  {van Albada} T.~S.,  2007, \mn@doi [\mnras]
  {10.1111/j.1365-2966.2007.11533.x}, \href
  {https://ui.adsabs.harvard.edu/abs/2007MNRAS.376.1513N} {376, 1513}

\bibitem[\protect\citeauthoryear{{Norris}, {Meidt}, {Van de Ven}, {Schinnerer},
  {Groves}  \& {Querejeta}}{{Norris} et~al.}{2014}]{Norris+2014}
{Norris} M.~A.,  {Meidt} S.,  {Van de Ven} G.,  {Schinnerer} E.,  {Groves} B.,
   {Querejeta} M.,  2014, \mn@doi [\apj] {10.1088/0004-637X/797/1/55}, \href
  {https://ui.adsabs.harvard.edu/abs/2014ApJ...797...55N} {797, 55}

\bibitem[\protect\citeauthoryear{{Obreschkow} \& {Glazebrook}}{{Obreschkow} \&
  {Glazebrook}}{2014}]{Obreschkow+2014}
{Obreschkow} D.,  {Glazebrook} K.,  2014, \mn@doi [\apj]
  {10.1088/0004-637X/784/1/26}, \href
  {https://ui.adsabs.harvard.edu/abs/2014ApJ...784...26O} {784, 26}

\bibitem[\protect\citeauthoryear{{Ogle}, {Lanz}, {Nader}  \& {Helou}}{{Ogle}
  et~al.}{2016}]{Ogle+2016}
{Ogle} P.~M.,  {Lanz} L.,  {Nader} C.,   {Helou} G.,  2016, \mn@doi [\apj]
  {10.3847/0004-637X/817/2/109}, \href
  {https://ui.adsabs.harvard.edu/abs/2016ApJ...817..109O} {817, 109}

\bibitem[\protect\citeauthoryear{{Ogle}, {Lanz}, {Appleton}, {Helou}  \&
  {Mazzarella}}{{Ogle} et~al.}{2019a}]{Ogle+2019a}
{Ogle} P.~M.,  {Lanz} L.,  {Appleton} P.~N.,  {Helou} G.,   {Mazzarella} J.,
  2019a, \mn@doi [\apjs] {10.3847/1538-4365/ab21c3}, \href
  {https://ui.adsabs.harvard.edu/abs/2019ApJS..243...14O} {243, 14}

\bibitem[\protect\citeauthoryear{{Ogle}, {Jarrett}, {Lanz}, {Cluver},
  {Alatalo}, {Appleton}  \& {Mazzarella}}{{Ogle} et~al.}{2019b}]{Ogle+2019}
{Ogle} P.~M.,  {Jarrett} T.,  {Lanz} L.,  {Cluver} M.,  {Alatalo} K.,
  {Appleton} P.~N.,   {Mazzarella} J.~M.,  2019b, \mn@doi [\apjl]
  {10.3847/2041-8213/ab459e}, \href
  {https://ui.adsabs.harvard.edu/abs/2019ApJ...884L..11O} {884, L11}

\bibitem[\protect\citeauthoryear{{Panter}, {Jimenez}, {Heavens}  \&
  {Charlot}}{{Panter} et~al.}{2007}]{Panter+2007}
{Panter} B.,  {Jimenez} R.,  {Heavens} A.~F.,   {Charlot} S.,  2007, \mn@doi
  [\mnras] {10.1111/j.1365-2966.2007.11909.x}, \href
  {https://ui.adsabs.harvard.edu/abs/2007MNRAS.378.1550P} {378, 1550}

\bibitem[\protect\citeauthoryear{{Parkash}, {Brown}, {Jarrett}  \&
  {Bonne}}{{Parkash} et~al.}{2018}]{Parkash+2018}
{Parkash} V.,  {Brown} M. J.~I.,  {Jarrett} T.~H.,   {Bonne} N.~J.,  2018,
  \mn@doi [\apj] {10.3847/1538-4357/aad3b9}, \href
  {https://ui.adsabs.harvard.edu/abs/2018ApJ...864...40P} {864, 40}

\bibitem[\protect\citeauthoryear{{Peebles}}{{Peebles}}{1969}]{Peebles+1969}
{Peebles} P.~J.~E.,  1969, \mn@doi [\apj] {10.1086/149876}, \href
  {https://ui.adsabs.harvard.edu/abs/1969ApJ...155..393P} {155, 393}

\bibitem[\protect\citeauthoryear{{Peletier} \& {Willner}}{{Peletier} \&
  {Willner}}{1993}]{Peletier+1993}
{Peletier} R.~F.,  {Willner} S.~P.,  1993, \mn@doi [\apj] {10.1086/173422},
  \href {https://ui.adsabs.harvard.edu/abs/1993ApJ...418..626P} {418, 626}

\bibitem[\protect\citeauthoryear{{Persic}, {Salucci}  \& {Stel}}{{Persic}
  et~al.}{1996}]{Persic+1996}
{Persic} M.,  {Salucci} P.,   {Stel} F.,  1996, \mn@doi [\mnras]
  {10.1093/mnras/278.1.27}, \href
  {https://ui.adsabs.harvard.edu/abs/1996MNRAS.281...27P} {281, 27}

\bibitem[\protect\citeauthoryear{{Planck Collaboration} et~al.,}{{Planck
  Collaboration} et~al.}{2016}]{PlanckCollaboration+2016}
{Planck Collaboration} et~al., 2016, \mn@doi [\aap]
  {10.1051/0004-6361/201525830}, \href
  {https://ui.adsabs.harvard.edu/abs/2016A&A...594A..13P} {594, A13}

\bibitem[\protect\citeauthoryear{{Ponomareva}, {Verheijen}, {Papastergis},
  {Bosma}  \& {Peletier}}{{Ponomareva} et~al.}{2018}]{Ponomareva+2018}
{Ponomareva} A.~A.,  {Verheijen} M. A.~W.,  {Papastergis} E.,  {Bosma} A.,
  {Peletier} R.~F.,  2018, \mn@doi [\mnras] {10.1093/mnras/stx3066}, \href
  {https://ui.adsabs.harvard.edu/abs/2018MNRAS.474.4366P} {474, 4366}

\bibitem[\protect\citeauthoryear{{Posti} \& {Fall}}{{Posti} \&
  {Fall}}{2021}]{PostiFall21}
{Posti} L.,  {Fall} S.~M.,  2021, \mn@doi [A\&A] {10.1051/0004-6361/202040256},
  \href {https://ui.adsabs.harvard.edu/abs/2021A&A...649A.119P} {649, A119}

\bibitem[\protect\citeauthoryear{{Posti}, {Pezzulli}, {Fraternali}  \& {Di
  Teodoro}}{{Posti} et~al.}{2018a}]{Posti+2018a}
{Posti} L.,  {Pezzulli} G.,  {Fraternali} F.,   {Di Teodoro} E.~M.,  2018a,
  \mn@doi [\mnras] {10.1093/mnras/stx3168}, \href
  {https://ui.adsabs.harvard.edu/abs/2018MNRAS.475..232P} {475, 232}

\bibitem[\protect\citeauthoryear{{Posti}, {Fraternali}, {Di Teodoro}  \&
  {Pezzulli}}{{Posti} et~al.}{2018b}]{Posti+2018b}
{Posti} L.,  {Fraternali} F.,  {Di Teodoro} E.~M.,   {Pezzulli} G.,  2018b,
  \mn@doi [\aap] {10.1051/0004-6361/201833091}, \href
  {https://ui.adsabs.harvard.edu/abs/2018A&A...612L...6P} {612, L6}

\bibitem[\protect\citeauthoryear{{Posti}, {Fraternali}  \& {Marasco}}{{Posti}
  et~al.}{2019a}]{Posti+2019b}
{Posti} L.,  {Fraternali} F.,   {Marasco} A.,  2019a, \mn@doi [\aap]
  {10.1051/0004-6361/201935553}, \href
  {https://ui.adsabs.harvard.edu/abs/2019A&A...626A..56P} {626, A56}

\bibitem[\protect\citeauthoryear{{Posti}, {Marasco}, {Fraternali}  \&
  {Famaey}}{{Posti} et~al.}{2019b}]{Posti+2019}
{Posti} L.,  {Marasco} A.,  {Fraternali} F.,   {Famaey} B.,  2019b, \mn@doi
  [\aap] {10.1051/0004-6361/201935982}, \href
  {https://ui.adsabs.harvard.edu/abs/2019A&A...629A..59P} {629, A59}

\bibitem[\protect\citeauthoryear{{Reyes}, {Mandelbaum}, {Gunn}, {Pizagno}  \&
  {Lackner}}{{Reyes} et~al.}{2011}]{Reyes+2011}
{Reyes} R.,  {Mandelbaum} R.,  {Gunn} J.~E.,  {Pizagno} J.,   {Lackner} C.~N.,
  2011, \mn@doi [\mnras] {10.1111/j.1365-2966.2011.19415.x}, \href
  {https://ui.adsabs.harvard.edu/abs/2011MNRAS.417.2347R} {417, 2347}

\bibitem[\protect\citeauthoryear{{R{\"o}ck}, {Vazdekis}, {Peletier}, {Knapen}
  \& {Falc{\'o}n-Barroso}}{{R{\"o}ck} et~al.}{2015}]{Rock+2015}
{R{\"o}ck} B.,  {Vazdekis} A.,  {Peletier} R.~F.,  {Knapen} J.~H.,
  {Falc{\'o}n-Barroso} J.,  2015, \mn@doi [\mnras] {10.1093/mnras/stv503},
  \href {https://ui.adsabs.harvard.edu/abs/2015MNRAS.449.2853R} {449, 2853}

\bibitem[\protect\citeauthoryear{{Romanowsky} \& {Fall}}{{Romanowsky} \&
  {Fall}}{2012}]{Romanowsky+2012}
{Romanowsky} A.~J.,  {Fall} S.~M.,  2012, \mn@doi [\apjs]
  {10.1088/0067-0049/203/2/17}, \href
  {https://ui.adsabs.harvard.edu/abs/2012ApJS..203...17R} {203, 17}

\bibitem[\protect\citeauthoryear{{Schmidt}}{{Schmidt}}{1959}]{Schmidt+1959}
{Schmidt} M.,  1959, \mn@doi [\apj] {10.1086/146614}, \href
  {https://ui.adsabs.harvard.edu/abs/1959ApJ...129..243S} {129, 243}

\bibitem[\protect\citeauthoryear{{Simard}, {Mendel}, {Patton}, {Ellison}  \&
  {McConnachie}}{{Simard} et~al.}{2011}]{Simard+2011}
{Simard} L.,  {Mendel} J.~T.,  {Patton} D.~R.,  {Ellison} S.~L.,
  {McConnachie} A.~W.,  2011, \mn@doi [\apjs] {10.1088/0067-0049/196/1/11},
  \href {https://ui.adsabs.harvard.edu/abs/2011ApJS..196...11S} {196, 11}

\bibitem[\protect\citeauthoryear{{Skrutskie} et~al.,}{{Skrutskie}
  et~al.}{2006}]{Skrutskie+2006}
{Skrutskie} M.~F.,  et~al., 2006, \mn@doi [\aj] {10.1086/498708}, \href
  {https://ui.adsabs.harvard.edu/abs/2006AJ....131.1163S} {131, 1163}

\bibitem[\protect\citeauthoryear{{Spekkens} \& {Giovanelli}}{{Spekkens} \&
  {Giovanelli}}{2006}]{Spekkens+2006}
{Spekkens} K.,  {Giovanelli} R.,  2006, \mn@doi [\aj] {10.1086/506177}, \href
  {https://ui.adsabs.harvard.edu/abs/2006AJ....132.1426S} {132, 1426}

\bibitem[\protect\citeauthoryear{{Torres-Flores}, {Epinat}, {Amram}, {Plana}
  \& {Mendes de Oliveira}}{{Torres-Flores} et~al.}{2011}]{Torres-Flores+2011}
{Torres-Flores} S.,  {Epinat} B.,  {Amram} P.,  {Plana} H.,   {Mendes de
  Oliveira} C.,  2011, \mn@doi [\mnras] {10.1111/j.1365-2966.2011.19169.x},
  \href {https://ui.adsabs.harvard.edu/abs/2011MNRAS.416.1936T} {416, 1936}

\bibitem[\protect\citeauthoryear{{Tully} \& {Fisher}}{{Tully} \&
  {Fisher}}{1977}]{Tully+1977}
{Tully} R.~B.,  {Fisher} J.~R.,  1977, \aap, \href
  {https://ui.adsabs.harvard.edu/abs/1977A&A....54..661T} {500, 105}

\bibitem[\protect\citeauthoryear{{Verheijen}}{{Verheijen}}{2001}]{Verheijen+2001}
{Verheijen} M. A.~W.,  2001, \mn@doi [\apj] {10.1086/323887}, \href
  {https://ui.adsabs.harvard.edu/abs/2001ApJ...563..694V} {563, 694}

\bibitem[\protect\citeauthoryear{{Wechsler} \& {Tinker}}{{Wechsler} \&
  {Tinker}}{2018}]{Wechsler+2018}
{Wechsler} R.~H.,  {Tinker} J.~L.,  2018, \mn@doi [\araa]
  {10.1146/annurev-astro-081817-051756}, \href
  {https://ui.adsabs.harvard.edu/abs/2018ARA&A..56..435W} {56, 435}

\bibitem[\protect\citeauthoryear{{Williams}, {Bureau}  \&
  {Cappellari}}{{Williams} et~al.}{2010}]{Williams+2010}
{Williams} M.~J.,  {Bureau} M.,   {Cappellari} M.,  2010, \mn@doi [\mnras]
  {10.1111/j.1365-2966.2010.17406.x}, \href
  {https://ui.adsabs.harvard.edu/abs/2010MNRAS.409.1330W} {409, 1330}

\bibitem[\protect\citeauthoryear{{Wright} et~al.,}{{Wright}
  et~al.}{2010}]{Wright+2010}
{Wright} E.~L.,  et~al., 2010, \mn@doi [\aj] {10.1088/0004-6256/140/6/1868},
  \href {https://ui.adsabs.harvard.edu/abs/2010AJ....140.1868W} {140, 1868}

\bibitem[\protect\citeauthoryear{{de los Reyes} \& {Kennicutt}}{{de los Reyes}
  \& {Kennicutt}}{2019}]{delosReyes+2019}
{de los Reyes} M. A.~C.,  {Kennicutt} Robert~C. J.,  2019, \mn@doi [\apj]
  {10.3847/1538-4357/aafa82}, \href
  {https://ui.adsabs.harvard.edu/abs/2019ApJ...872...16D} {872, 16}

\bibitem[\protect\citeauthoryear{{den Heijer} et~al.,}{{den Heijer}
  et~al.}{2015}]{denHeijer+2015}
{den Heijer} M.,  et~al., 2015, \mn@doi [\aap] {10.1051/0004-6361/201526879},
  \href {https://ui.adsabs.harvard.edu/abs/2015A&A...581A..98D} {581, A98}

\bibitem[\protect\citeauthoryear{{van Albada}, {Bahcall}, {Begeman}  \&
  {Sancisi}}{{van Albada} et~al.}{1985}]{vanAlbada+1985}
{van Albada} T.~S.,  {Bahcall} J.~N.,  {Begeman} K.,   {Sancisi} R.,  1985,
  \mn@doi [\apj] {10.1086/163375}, \href
  {https://ui.adsabs.harvard.edu/abs/1985ApJ...295..305V} {295, 305}

\makeatother
\end{thebibliography}

\newpage 

\begin{appendix}
\section{Atlas of extremely massive spiral galaxies}
\label{appendixA}
In \autoref{fig:appfig} of this Appendix, we show surface-brightness profiles, kinematic models and rotation curves derived for all \galnum\ massive discs in our sample. For each galaxy we plot, from left to right: 

\begin{itemize}[itemsep=2pt,leftmargin=*]
    \item[1)] RGB image in the $g$, $r$ and $z$ bands from the Legacy surveys \citep{Dey+2019}.
    
    \item[2)] Surface-brigthness profiles derived from $z$-band images (Legacy survey, blue dots) and from W1-band images (WISE, cyan triangles) through elliptical isophote fitting (see Sections~\ref{sec:profiles}). We also plot our best-fit model with a double Sérsic function of the W1-band profile: the dashed red line denotes the bulge component, the dotted-dashed green line the disk component and the black line the total model. Fits are only made outside of the PSF, which is 5$''$ for the WISE W1 band.
    
    \item[3)] Long-slit \ha\ emission-line data (grey scale and contours) and best-fit kinematic model (red contours) described in Section~\ref{sec:kinmod}.
    Contour levels are at $2^n \times 2.5\sigma_\mathrm{rms}$, where $\sigma_\mathrm{rms}$ is the rms noise of the data and $n=0,1,2,3,4$. Yellow dots are the best-fit rotation velocities projected along the line of sight (i.e.\ $\vrot\sin i$).
    
    \item[4)] Best-fit rotation curve (red points). Empty dots denote regions where the rotation velocity is not well determined because of possible projection/absorption effects (see Section~\ref{sec:kinmod}). Errors in the rotation velocities take into account systematic uncertainties on the inclination angle and long-slit positioning.
    The velocity of the flat part of the rotation curve $\vflat$, used in the Tully-Fisher relations, and its error are shown as a grey horizontal line and a grey band, respectively.
\end{itemize}

\begin{figure*}
	\includegraphics[width=1\textwidth]{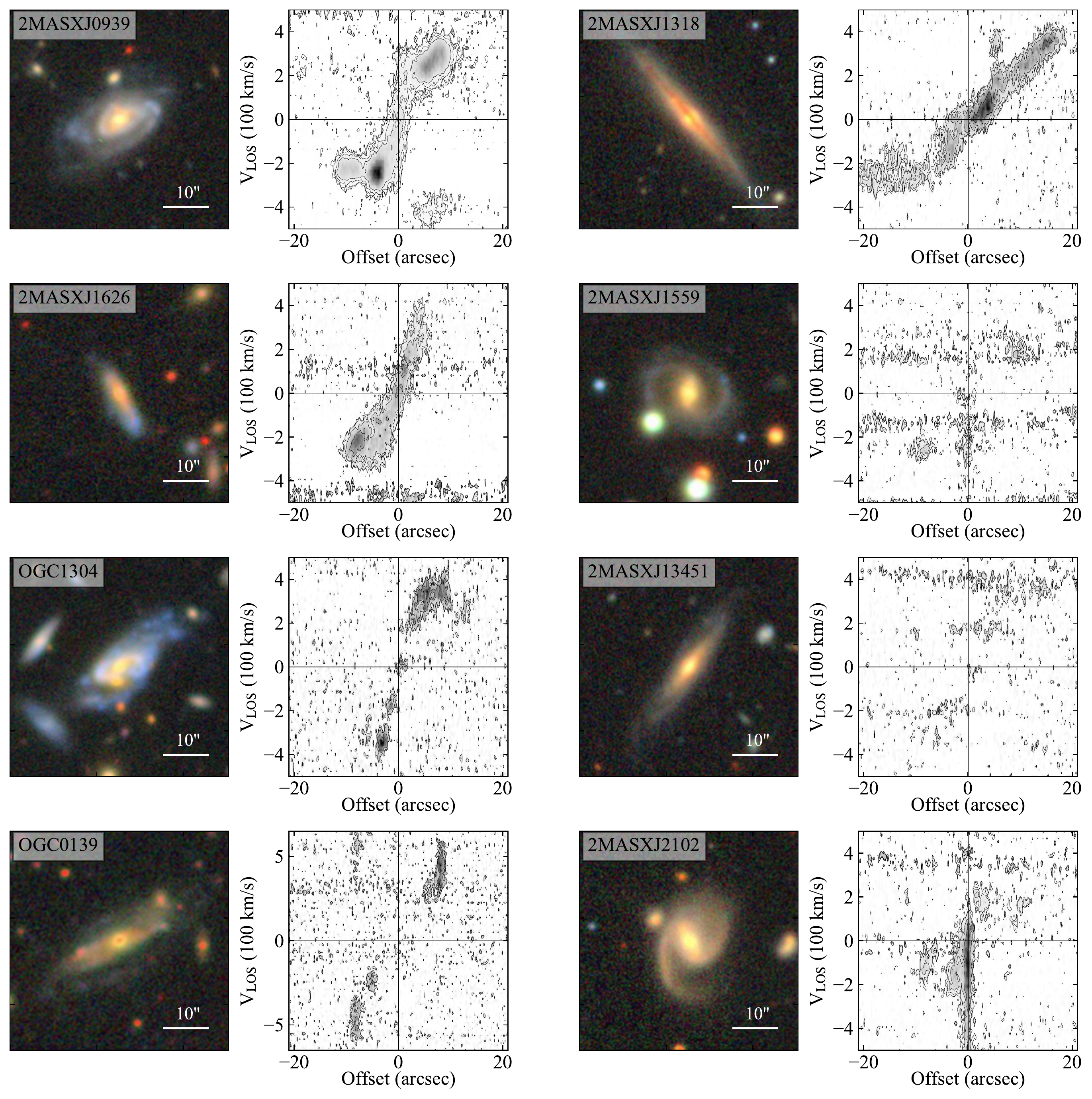}
    \caption{The 8 galaxies excluded from our kinematic analysis.}
    \label{fig:discarded}
\end{figure*}

\noindent \autoref{fig:discarded} shows composite $g,r,z$ images (color) from the Legacy and \ha\ long-slit SALT data (grey scale) for the 6 galaxies that were excluded from our kinematic analysis because we could not derive unambiguous rotation curves. Contours are at the same levels as in \autoref{fig:appfig}.

\begin{figure*}
	\includegraphics[width=1\textwidth]{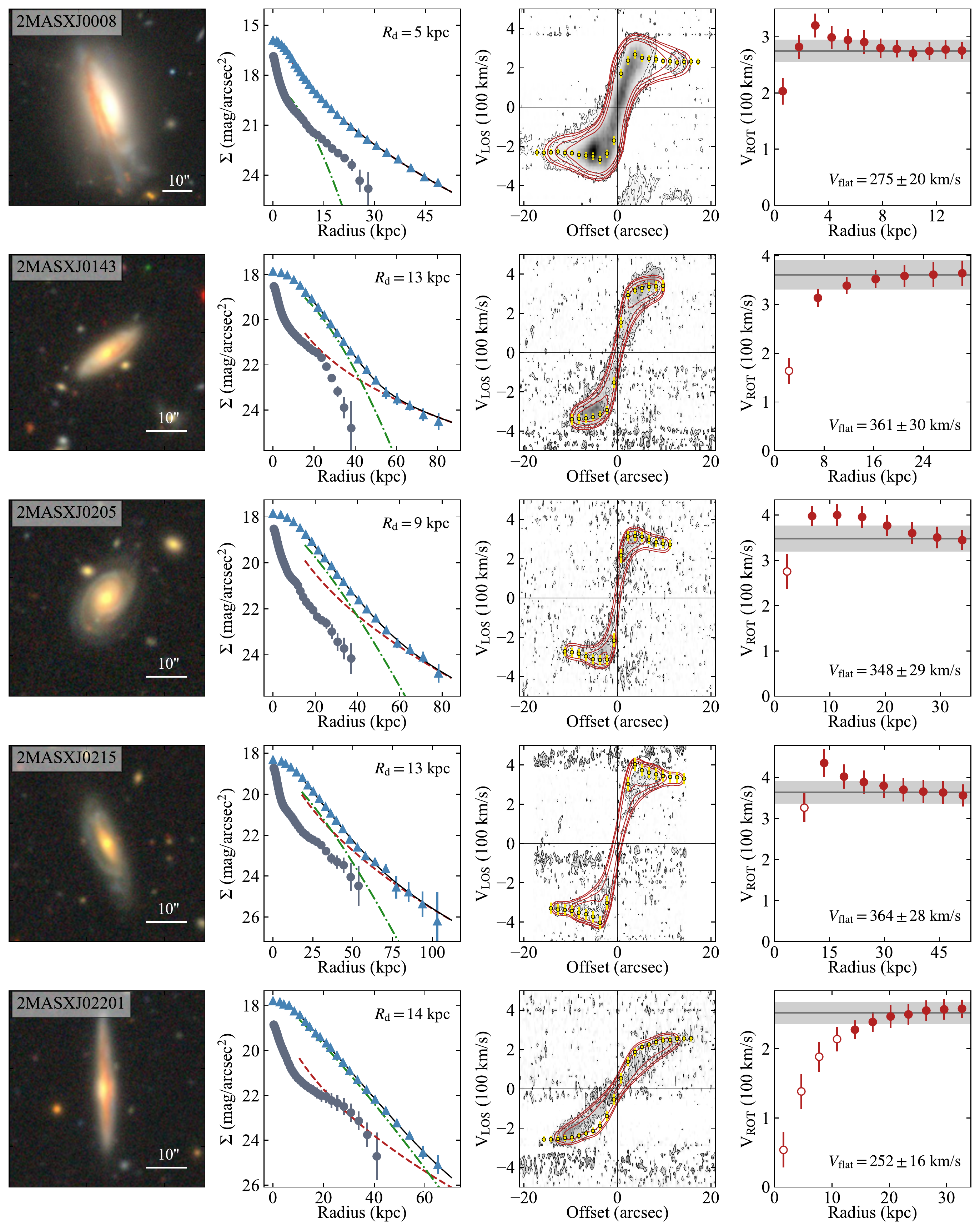}
    \caption{Surface photometry and kinematic analysis of \galnum\ massive spiral galaxies. See Appendix~\ref{appendixA} for details on the plots.}
    \label{fig:appfig}
\end{figure*}
\begin{figure*}\ContinuedFloat
	\includegraphics[width=1\textwidth]{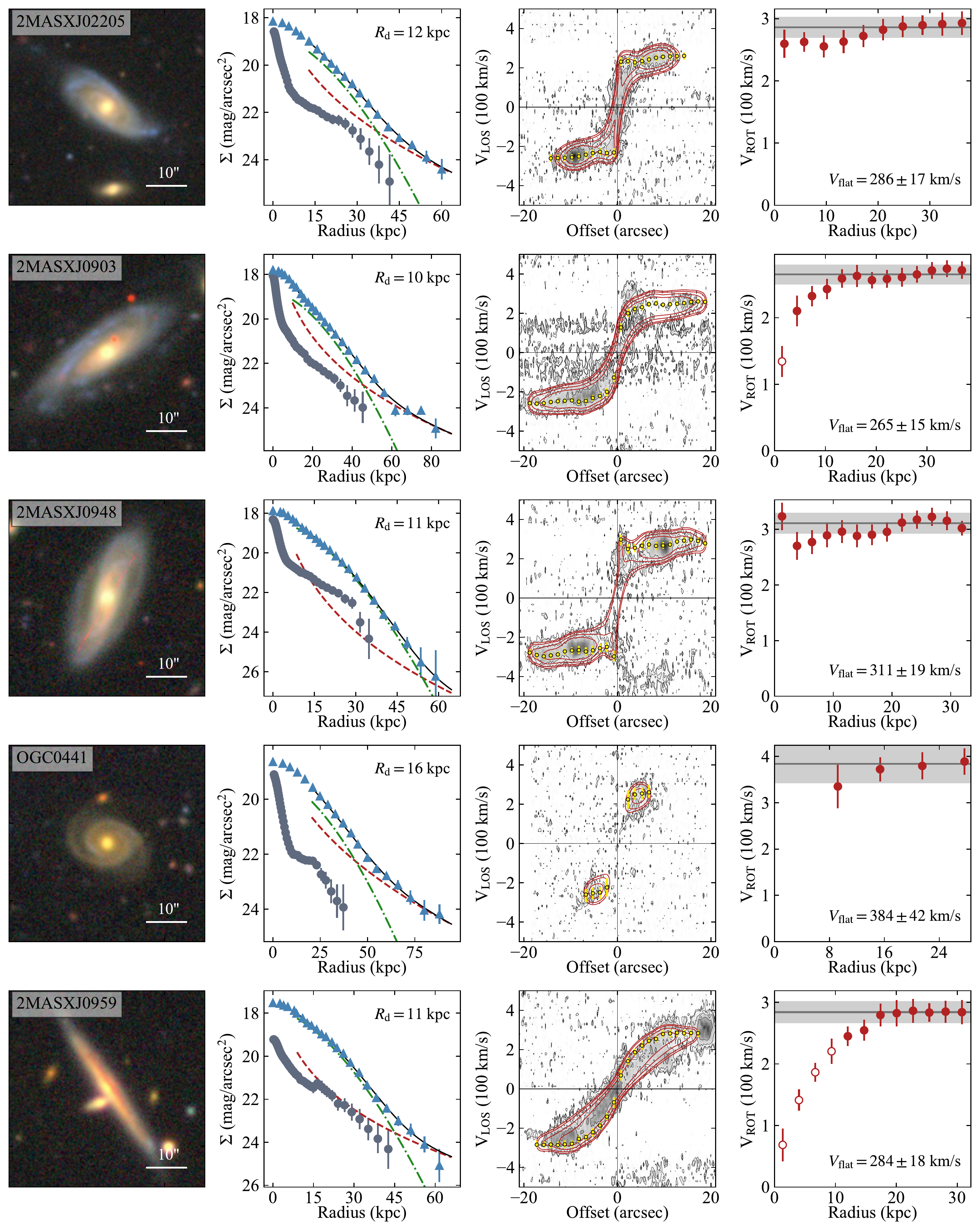}
    \caption{Continued}
\end{figure*}
\begin{figure*}\ContinuedFloat
	\includegraphics[width=1\textwidth]{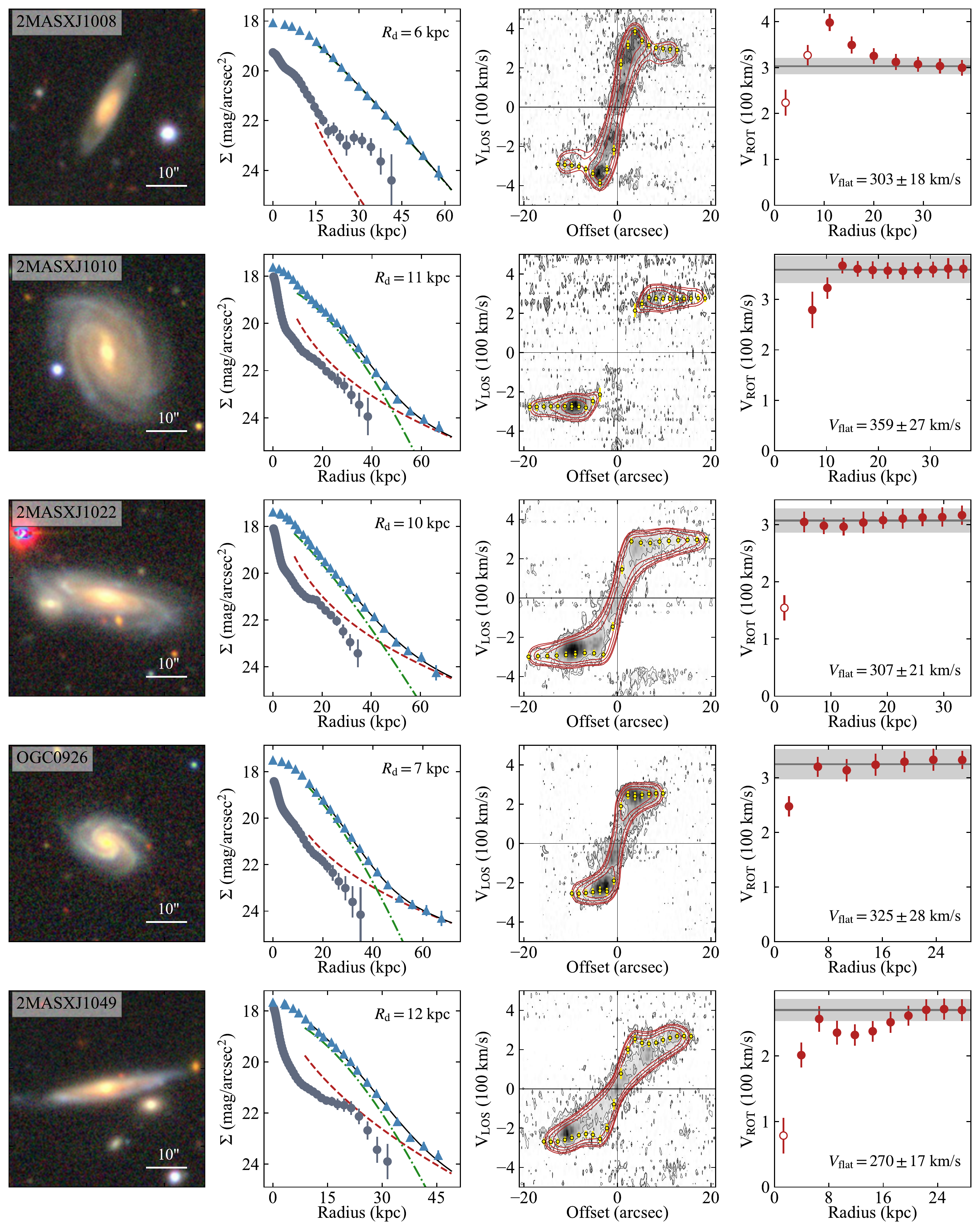}
    \caption{Continued}
\end{figure*}
\begin{figure*}\ContinuedFloat
	\includegraphics[width=1\textwidth]{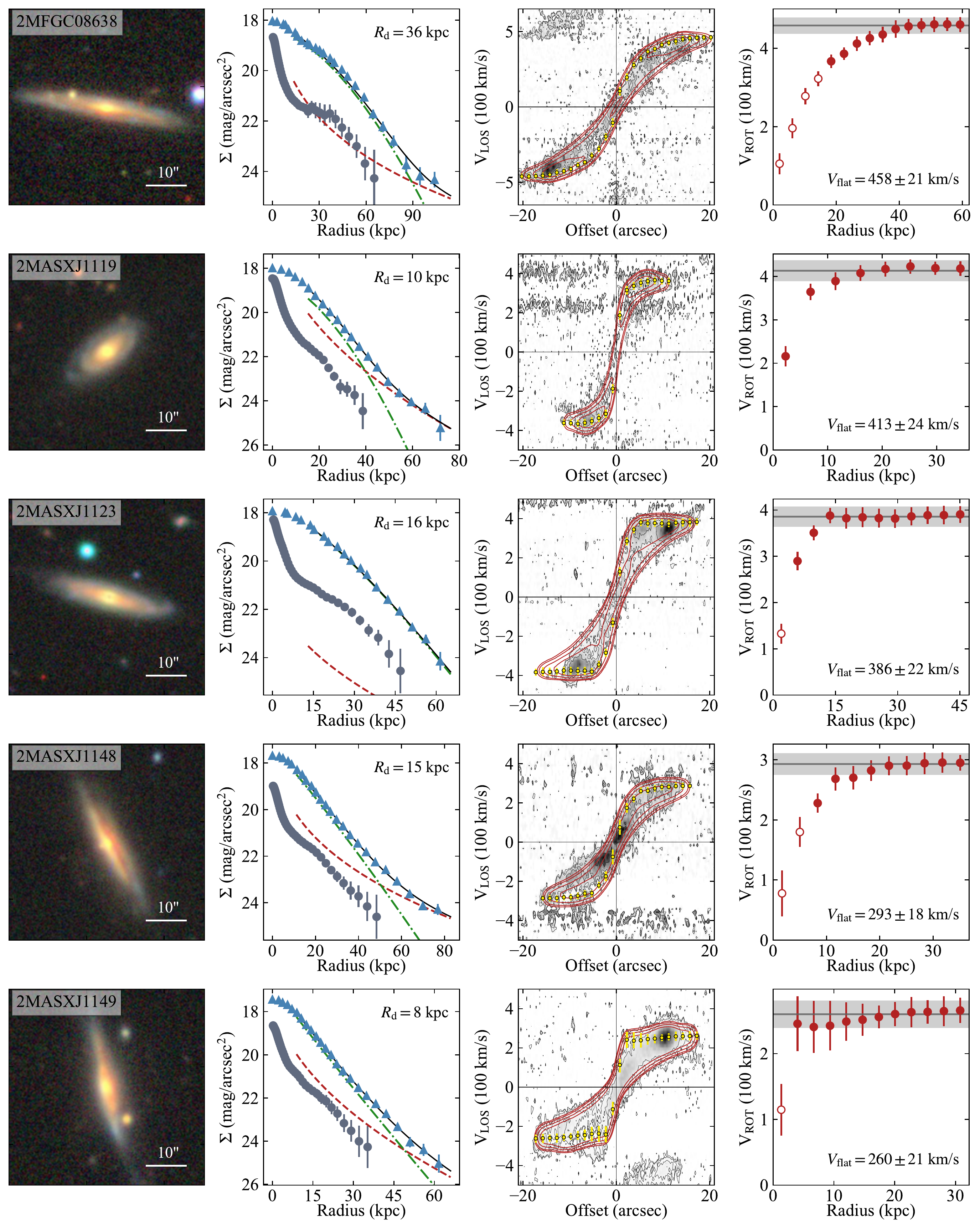}
    \caption{Continued}
\end{figure*}
\begin{figure*}\ContinuedFloat
	\includegraphics[width=1\textwidth]{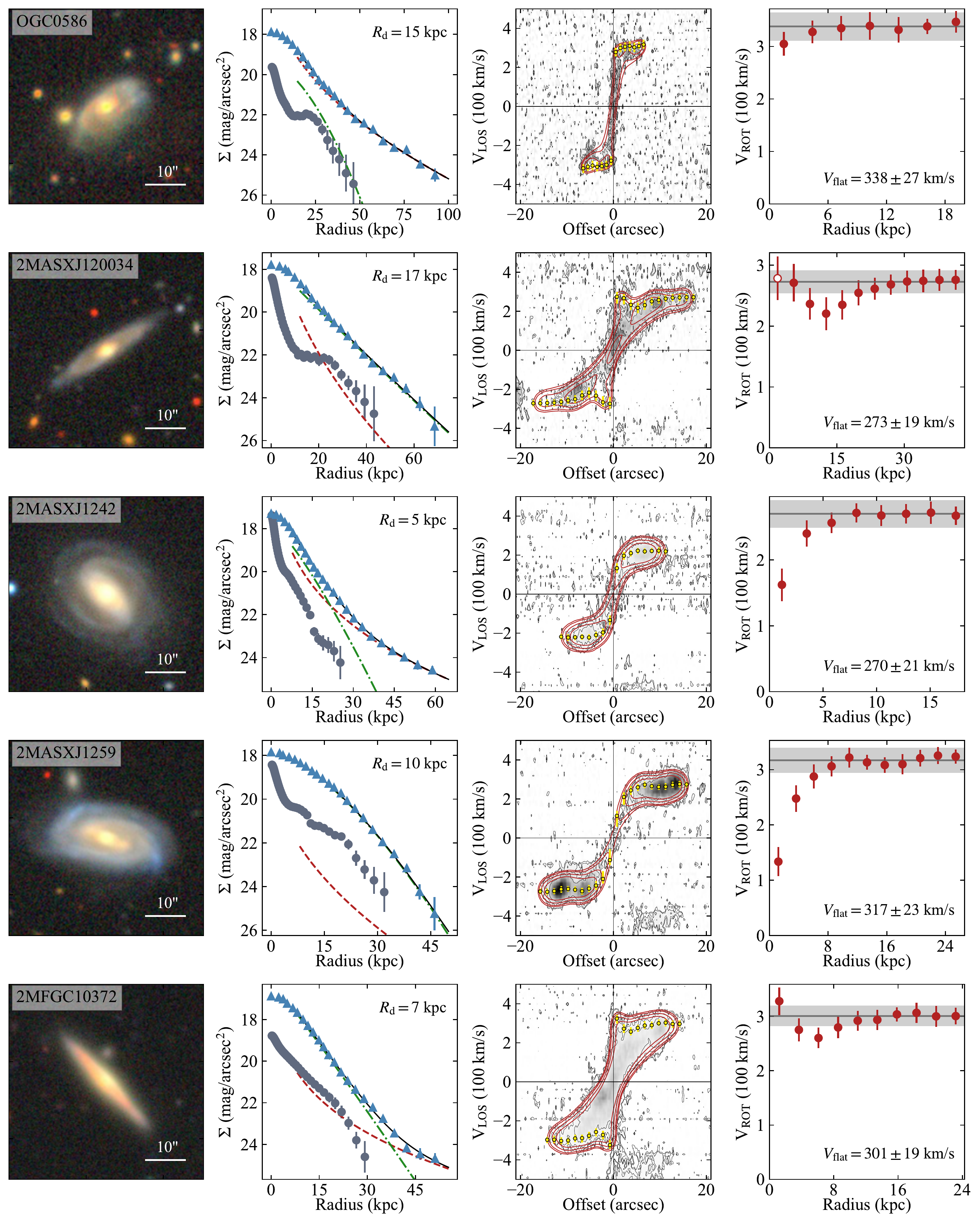}
    \caption{Continued}
\end{figure*}
\begin{figure*}\ContinuedFloat
	\includegraphics[width=1\textwidth]{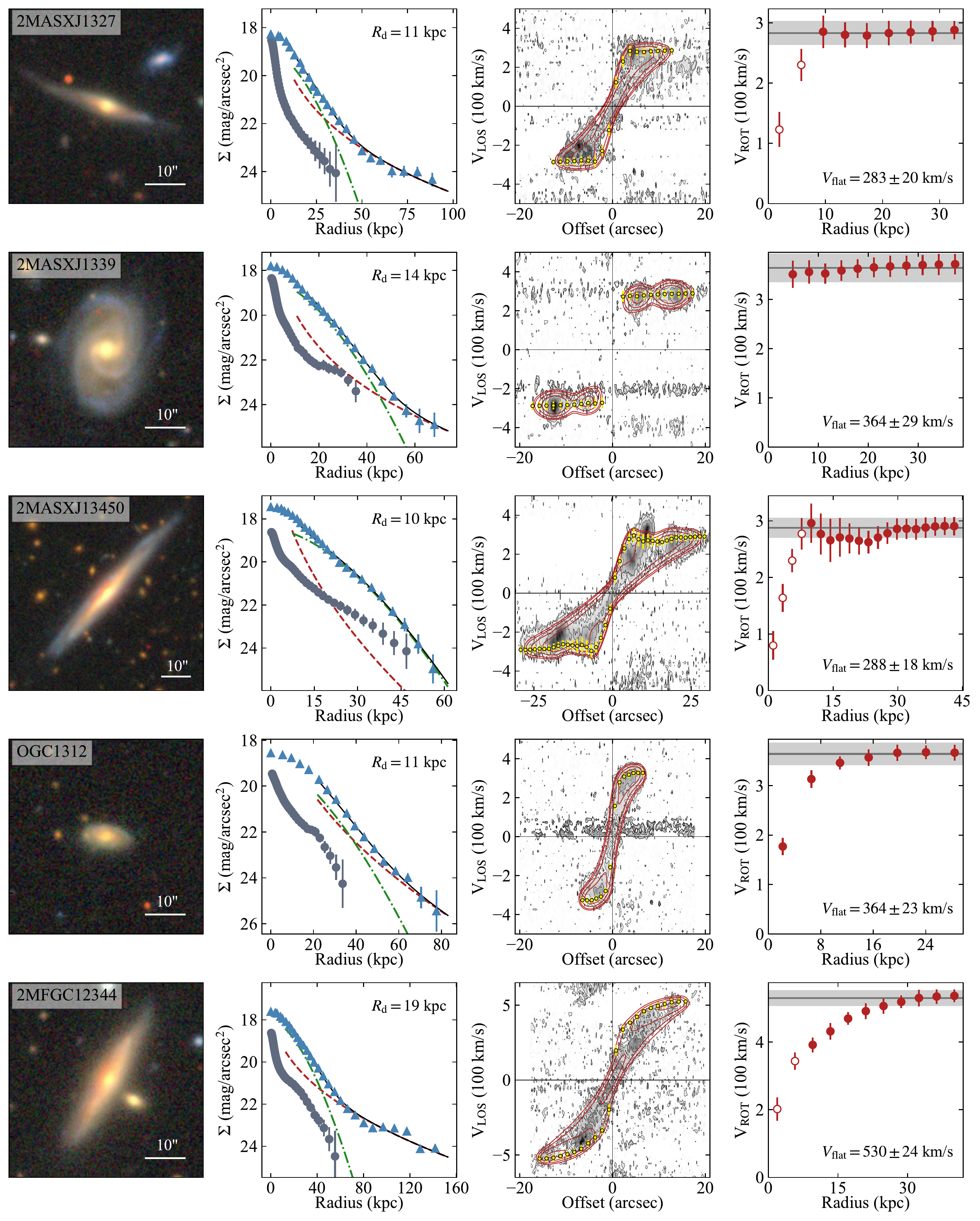}
    \caption{Continued}
\end{figure*}
\begin{figure*}\ContinuedFloat
	\includegraphics[width=1\textwidth]{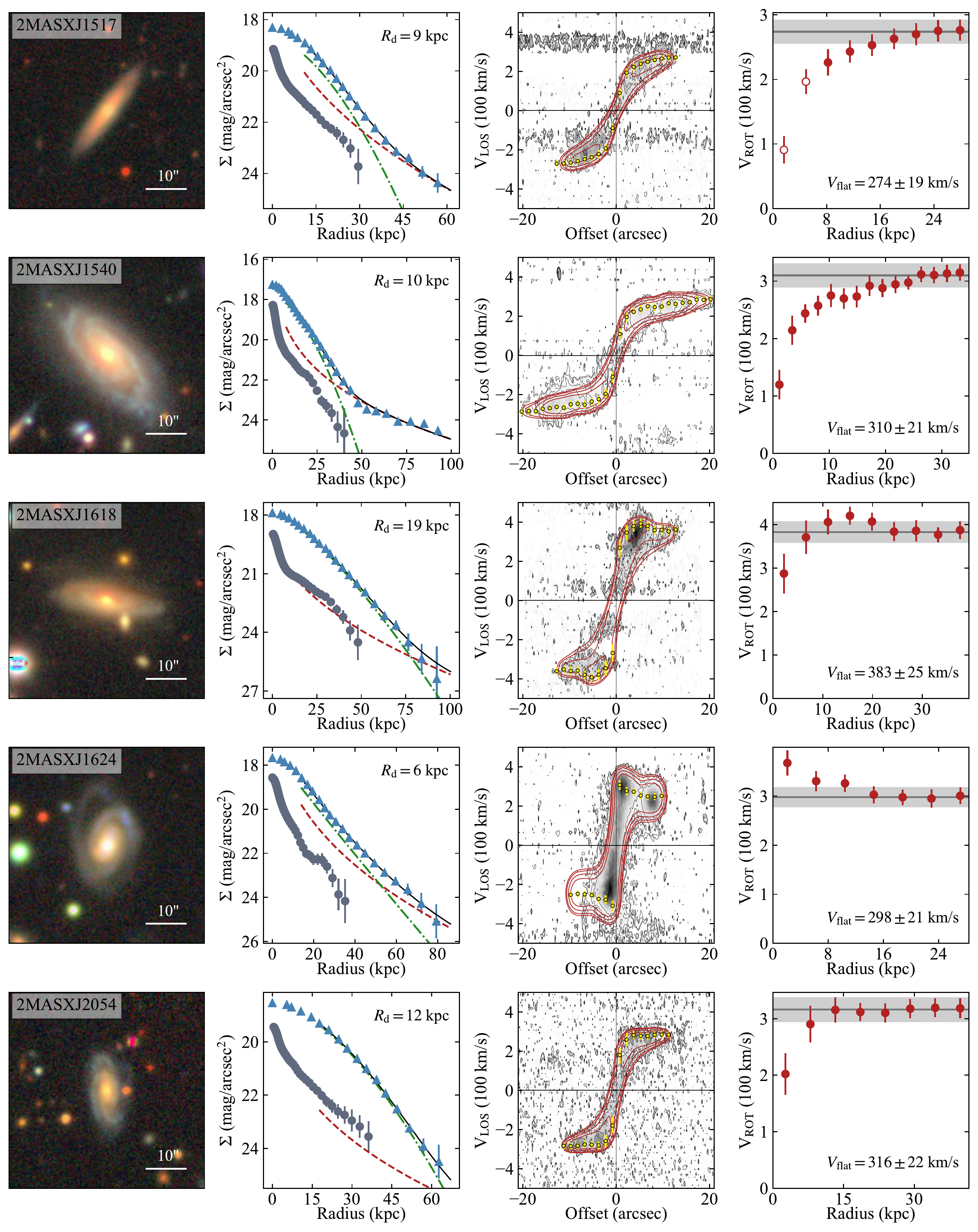}
    \caption{Continued}
\end{figure*}
\begin{figure*}\ContinuedFloat
	\includegraphics[width=1\textwidth]{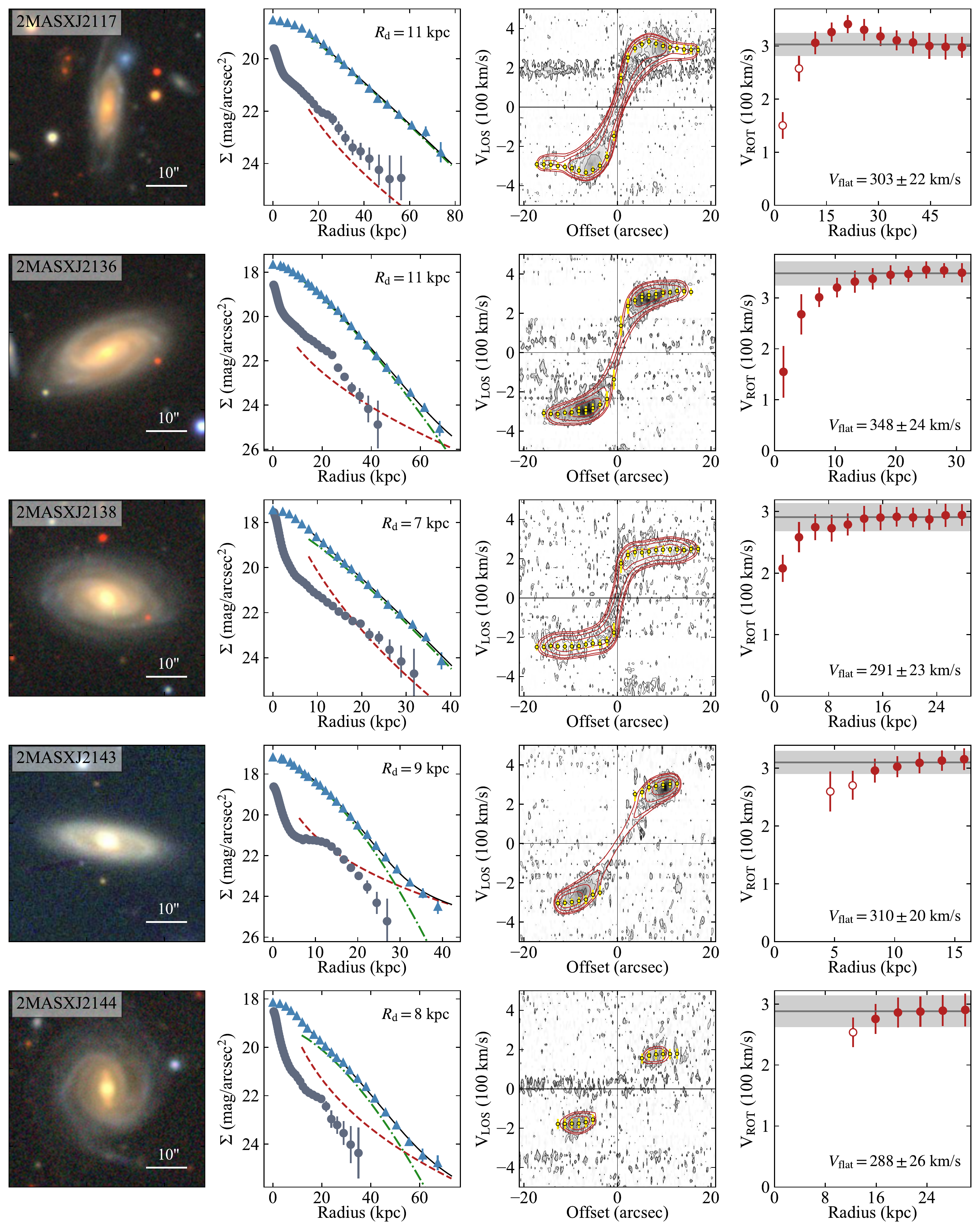}
    \caption{Continued}
\end{figure*}
\begin{figure*}\ContinuedFloat
	\includegraphics[width=1\textwidth]{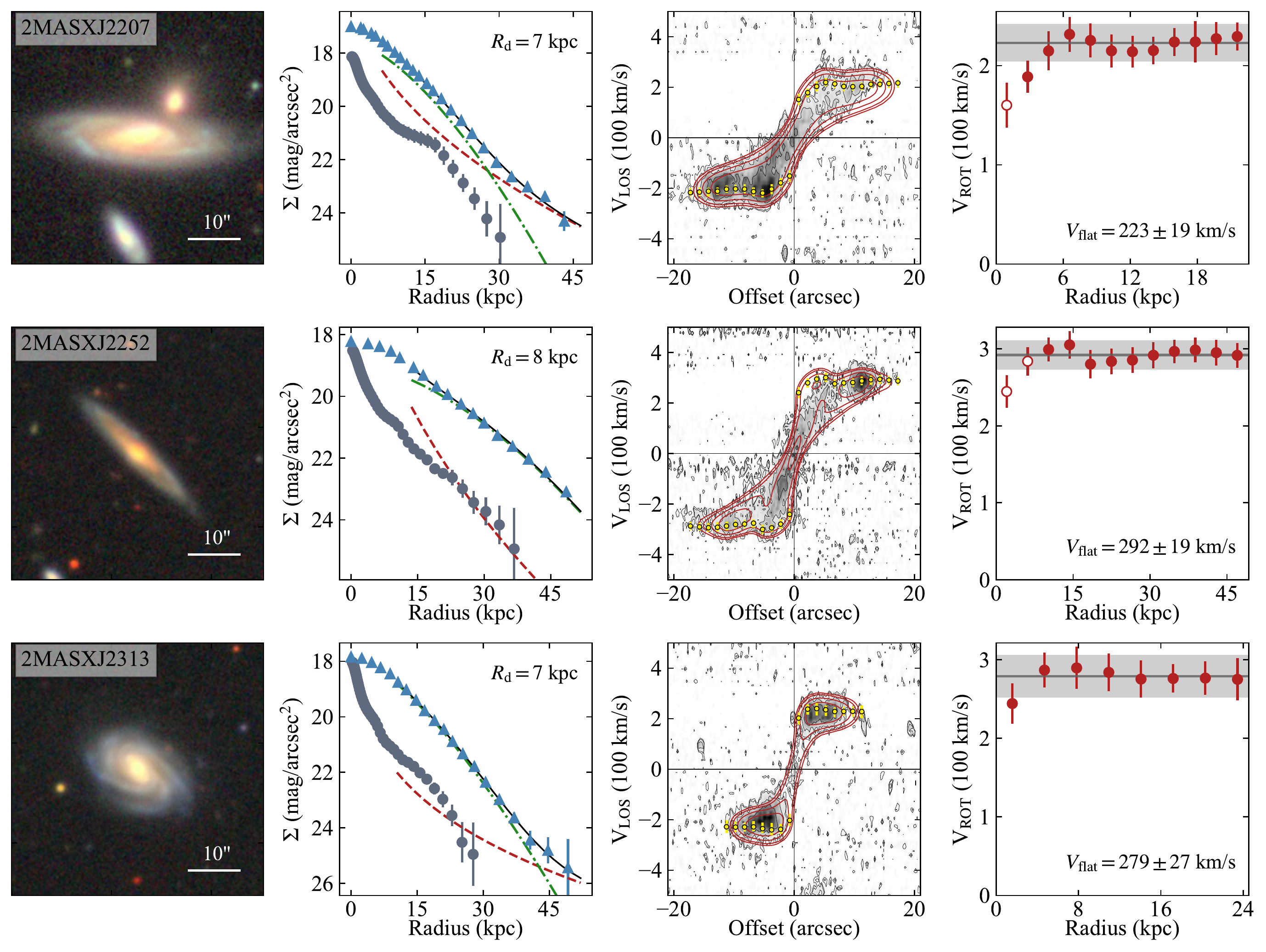}
    \caption{Continued}
\end{figure*}

\end{appendix}

\label{lastpage}
\end{document}